\pdfoutput=1
\documentclass[a4paper,11pt]{article}
\usepackage{amsmath,amssymb,amsfonts}
\usepackage[multiple,stable]{footmisc}
\usepackage[amsmath,amsthm,thmmarks]{ntheorem}
\allowdisplaybreaks[4]
\usepackage[noconfig]{refstyle}
\usepackage[dvipsnames]{xcolor}
\usepackage[hyperfootnotes=false, linktocpage=true, colorlinks, citecolor=blue, linkcolor=blue, urlcolor=Maroon]{hyperref}
\usepackage{array,tabularx,booktabs,geometry,subfigure,graphicx,longtable}
\usepackage[bf]{caption}
\usepackage{tikz}
\usetikzlibrary{shapes,arrows}
\tikzstyle{block} = [rectangle, draw, text width=7em, text centered, rounded corners, minimum height=3em]
\usepackage{graphicx,color}
\usepackage{cite, setspace, bigstrut, framed, eufrak, pifont}
\usepackage{wasysym}
\usepackage{arydshln}
\usepackage{float}

\newcommand{\eref}[1]{(\ref{#1})}

\newcommand{\eeq}{\end{equation}}
\newcommand{\beq}{\begin{equation}}
\newcommand{\eea}{\end{eqnarray}}
\newcommand{\bea}{\begin{eqnarray}}
\newcommand{\ba}{\begin{array}}
\newcommand{\ea}{\end{array}}
\newcommand{\bi}{\begin{itemize}}
\newcommand{\ei}{\end{itemize}}
\newcommand{\cL}{{\cal L}}

\newcommand{\nn}{\nonumber}

\newcommand{\cO}{{\cal O}}
\newcommand{\cA}{{\cal A}}
\newcommand{\IP}{\mathbb P}
\newcommand{\IC}{\mathbb C}
\newcommand{\IZ}{\mathbb Z}

\newcommand{\be}{\begin{equation}}
\newcommand{\ee}{\end{equation}}

\newcommand{\iddots}{\mathinner{\mkern2mu\raise1pt\hbox{.}\mkern2mu \raise4pt\hbox{.}\mkern2mu\raise7pt\hbox{.}\mkern1mu}}

\providecommand{\id}{\leavevmode\hbox{\small$\mathrm{1}$\kern-3.8pt\normalsize$\mathrm{1}$}}
\def\fnote#1#2{\begingroup\def\thefootnote{#1}\footnote{#2}
     \addtocounter{footnote}{-1}\endgroup}

\makeatletter \@addtoreset{equation}{section} \makeatother

\begin{document}

\vspace{1cm}

\title{
       \vskip 40pt
       {\huge \bf Tools for CICYs in F-theory}}

\vspace{2cm}

\author{Lara B. Anderson${}^{}$, Xin Gao${}^{}$, James Gray${}^{}$ and Seung-Joo Lee${}^{}$}
\date{}
\maketitle
\begin{center} {\small ${}^{}${\it Physics Department, Robeson Hall, Virginia Tech, Blacksburg, VA 24061, USA}}
\fnote{}{lara.anderson@vt.edu, xingao@vt.edu, jamesgray@vt.edu, seungsm@vt.edu}
\end{center}

\begin{abstract}
\noindent  We provide a set of tools for analyzing the geometry of elliptically fibered Calabi-Yau manifolds, starting with a description of the total space rather than with a Weierstrass model or a specified type of fiber/base. Such an approach to the subject of F-theory compactification makes certain geometric properties, which are usually hidden, manifest. Specifically, we review how to isolate genus-one fibrations in such geometries and then describe how to find their sections explicitly. This includes a full parameterization of the Mordell-Weil group where non-trivial. We then describe how to analyze the associated Weierstrass models, Jacobians and resolved geometries. We illustrate our discussion with concrete examples which are complete intersections in products of projective spaces (CICYs). The examples presented include cases exhibiting non-abelian symmetries and higher rank Mordell-Weil group. We also make some comments on non-flat fibrations in this context. In a companion paper \cite{phys} to this one, these results will be used to analyze the consequences for string dualities of the ubiquity of multiple fibrations in known constructions of Calabi-Yau manifolds.
\end{abstract}

\thispagestyle{empty}
\setcounter{page}{0}
\newpage

\tableofcontents

\newpage

\section{Introduction}

In many approaches to compactifications of F-theory, the identification of the fiber and base of the internal manifold is built in from the start. Typically, one begins with a choice of base manifold, and then fibers an elliptic curve, described in terms of an appropriate complete intersection in some toric variety, over that space. Systematic scans over Calabi-Yau (CY) geometries constructed in this manner can be found in Refs.~\cite{Braun:2011ux,Morrison:2012np,Morrison:2012js,Taylor:2012dr,Martini:2014iza,Anderson:2014gla,Johnson:2014xpa,Taylor:2015isa,Halverson:2015jua,Taylor:2015ppa,Johnson:2016qar}. Such a construction has many advantages, including the fact that one is guaranteed that the associated manifold is genus-one fibered and therefore is indeed suitable for use in F-theory. Nevertheless, this methodology has the drawback that, instead of simply using the large data sets of Calabi-Yau manifolds that have already been constructed (for example in Refs.~\cite{Yau:1986gu,Hubsch:1986ny,Candelas:1987kf,Candelas:1987du,Green:1986ck,Gray:2013mja,Gray:2014fla,Kreuzer:2000xy,Kreuzer:2002uu,Rohsiepe:2005qg,Altman:2014bfa}), one is essentially starting all over again in reconstructing those manifolds with the desired internal structure. In addition, as we will discuss, certain properties of the resulting compactifications can be hard to see in such descriptions.

In this paper we present tools for systematically pursuing a different approach to F-theory compactification. We describe how to take any smooth Calabi-Yau manifold and extract the F-theory physics associated to this ``resolved space" directly (for some related work see Ref.~\cite{Rohsiepe:2005qg}). In particular we describe how to do the following:
\begin{enumerate}
\item Isolate genus-one fibrations in one of the conventional data sets of Calabi-Yau manifolds. Here we build on technology first developed in Refs.~\cite{Kreuzer:1997zg,Rohsiepe:2005qg,Braun:2011ux,Gray:2014fla}.
\item Determine whether or not each of these fibrations have a section. In cases where a section is present we describe how to obtain an explicit form for it in terms of the original description of the manifold. This methodology is closely related to descriptions of holomorphic functions used in recent constructions of ``generalized complete intersection CY manifolds" (gCICYs) \cite{Anderson:2015iia}.
\item Obtain an explicit Weierstrass model associated to blowing down all components of the fibers in the original manifold that have generic vanishing intersection number with a chosen zero section. Here we follow the construction of Refs.~\cite{deligne,nakayama}.
\item Obtain the Jacobian manifold associated to the original compactification, making use of techniques from \cite{art, Braun:2011ux, Braun:2014qka}.
\end{enumerate}

Once this data has been obtained, standard techniques can be employed to study the F-theoretical physics of the compactifications in question. We illustrate all of this with concrete examples taken from the data set of complete intersection Calabi-Yau (CICY) manifolds~\cite{Yau:1986gu,Hubsch:1986ny,Green:1986ck,Candelas:1987kf,Candelas:1987du,Gray:2013mja,Gray:2014fla}. We expect that similar techniques should allow the methodology to be generalized to other data sets such as that due to Kreuzer and Skarke \cite{Kreuzer:2000xy,Kreuzer:2002uu,Rohsiepe:2005qg,Altman:2014bfa}, or the gCICYs \cite{Anderson:2015iia,Anderson:2015yzz,Berglund:2016yqo}, in a straightforward manner. Note that some steps in this direction have already been taken with Jacobian forms associated to complete intersection fibers Ref.~\cite{Braun:2014qka}.

\begin{figure}[!tbp]
  \centering
  \begin{minipage}[b]{0.45\textwidth}
    \includegraphics[scale=0.55]{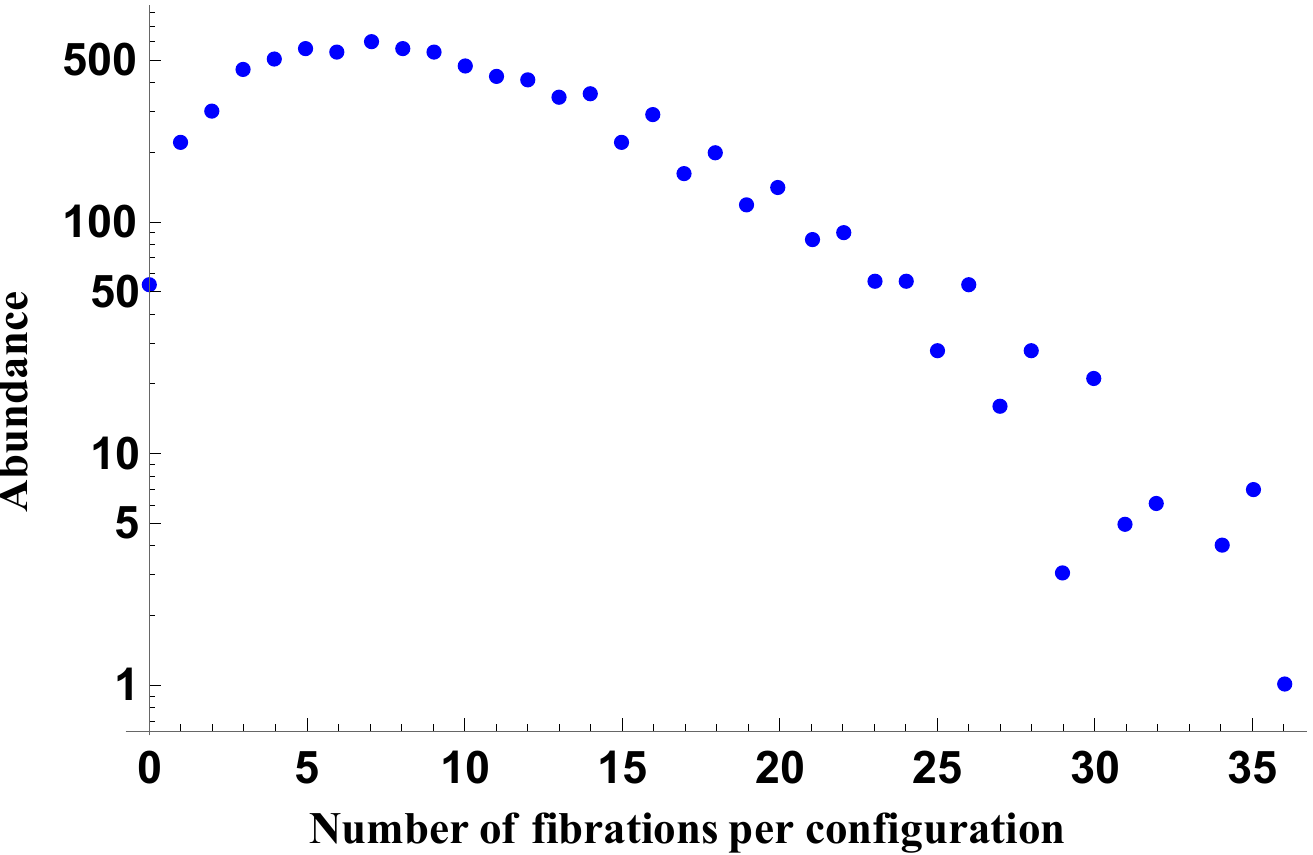}
\caption{\footnotesize  The abundance of CICY threefold configuration matrices, in the standard list, exhibiting a given number of genus-one fibrations which are visible directly in the configuration matrix \cite{usscanning}.}\label{fibrations1}
  \end{minipage}
  \hspace{0.1cm}
  \begin{minipage}[b]{0.45\textwidth}
   \includegraphics[scale=0.3]{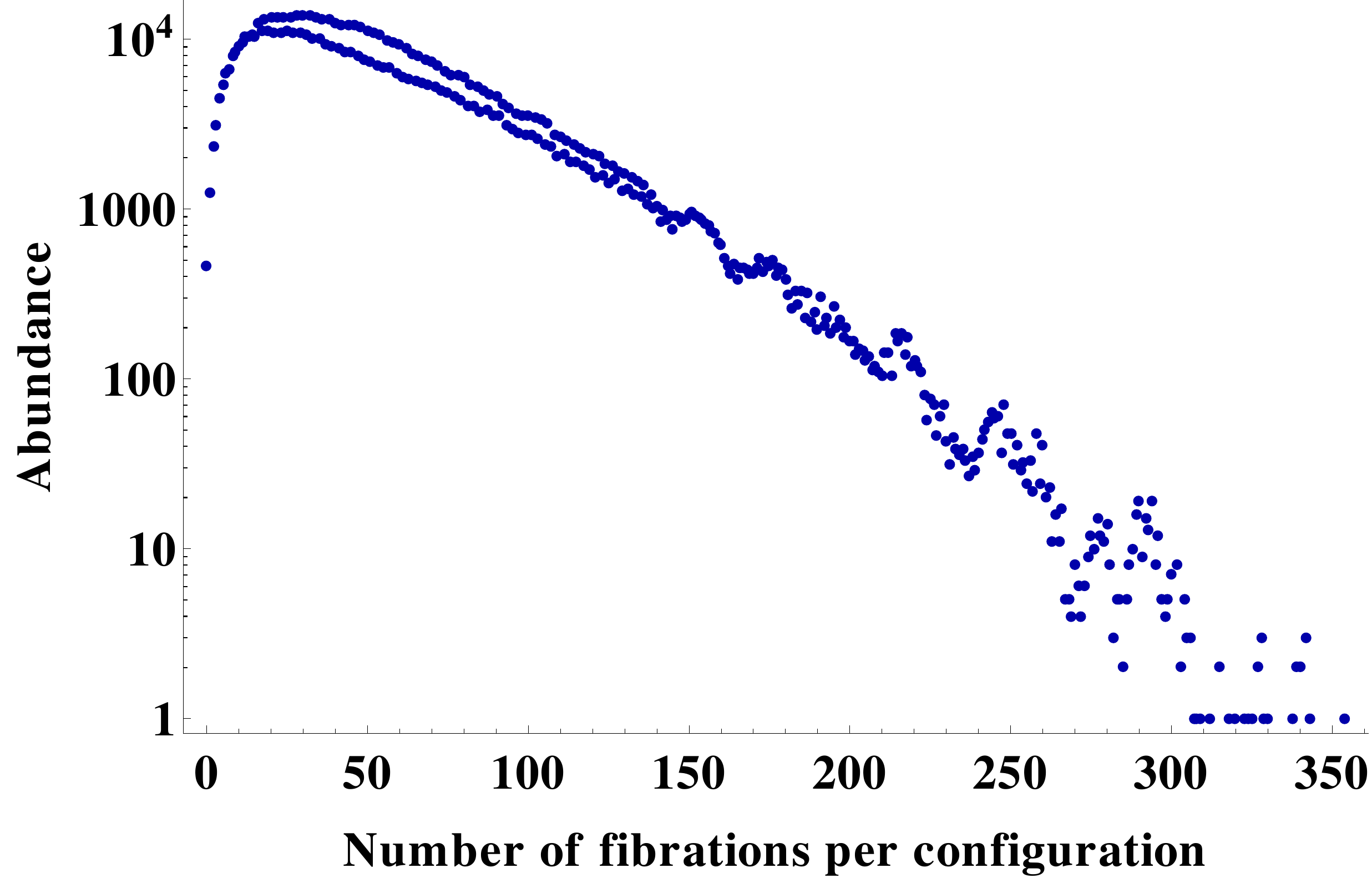}
\caption{\footnotesize  The abundance of CICY fourfold configuration matrices, in the standard list, exhibiting a given number of genus-one fibrations which are visible directly in the configuration matrix \cite{Gray:2014fla}.}\label{fibrations2}
  \end{minipage}
\end{figure}

In addition to utilizing pre-existing data sets of Calabi-Yau manifolds, this approach to analyzing global F-theory compactifications makes evident some features of fibrations that are not as obvious in more standard methodologies. One of the most important of these features, which will be explored extensively in the context of dualities in a companion paper \cite{phys} to this one, is that of multiple fibrations in a single CY geometry.
It is known \cite{Rohsiepe:2005qg,Johnson:2014xpa,Gray:2014fla,Johnson:2016qar} that the vast majority of known Calabi-Yau manifolds are genus-one fibered. It is also suspected that essentially all such manifolds can be written in a myriad of different ways as such a fibration -- indeed this has been proven in the case of complete intersections in products of projective spaces \cite{Gray:2014fla,usscanning}. This abundance of possible rewritings of the CICY geometries is illustrated in Figures \ref{fibrations1} and \ref{fibrations2}. 
If one simply constructs a Calabi-Yau manifold as a particular fiber type over a given base (e.g. in Weierstrass form), the existence of other descriptions of the manifold as fibrations over different bases can be difficult to see. In the approach we are following here, such collections of descriptions are manifest, and thus the obviously closely related F-theory models that they correspond to can be studied as a set \cite{Rohsiepe:2005qg,phys}.

As a simple illustration of multiple genus one fibrations, consider the following Calabi-Yau threefold, described as a complete intersection of two polynomial equations in $\IP^1 \times \IP^2 \times \IP^1 \times \IP^1$. 
\beq \label{intro1}
\quad X =\def\arraystretch{1.2}\left[\ba{c||cc} 
\IP^1 &  1 & 1 \\
\IP^2 &  1 & 2 \\ 	
\IP^1 &  1 & 1 \\
\IP^1 & 1 & 1 \\ 
\ea\right]  \ ,  
\eeq
Here the columns describe the two defining relations of the Calabi-Yau manifold by giving their polynomial degree in the homogeneous coordinates of the ambient projective spaces. This construction is described in detail in the next section. The manifold $X$ can be written in two different ways as a genus-one fibration. Below the fiber is described by the configuration matrix above the dotted line:
\beq
\def\arraystretch{1.2}\left[\ba{c||cc} 
\IP^1 &  1 & 1 \\
\IP^2 &  1 & 2 \\ \hdashline
\IP^1 &  1 & 1 \\
\IP^1 & 1 & 1 \\ 
\ea\right]   \;\;\ , \;\;\;\;\;\;\; 
\def\arraystretch{1.2}\left[\ba{c||cc} 
\IP^1 &  1 & 1 \\
\IP^1 & 1 & 1 \\ 
\IP^1 &  1 & 1 \\ \hdashline
\IP^2 &  1 & 2 \\
\ea\right]  \ .
\eeq
The base of the fibration is then simply $\IP^1 \times \IP^1$ in the first case and $\IP^2$ in the second.

For the first of these fibrations we will provide, in Section~\ref{eg2}, an explicit description of the associated Weierstrass model (over $\IP^1 \times \IP^1$) as one simple example of our method. The second of these fibrations is genus one but does not have a section. We provide several examples in Sections \ref{eg1}-\ref{Sec:X3-3} with some exhibiting multiple elliptic fibrations giving rise to the same total space.

A second feature that we have observed in applying this approach to the CICY data set, is that a great deal of these manifolds exhibit at least one fibration with a relatively high rank Mordell-Weil group. Indeed, even for the very simple example given above, the first fibration discussed admits two sections. As will be detailed in Section~\ref{eg2}, these can be described as the global holomorphic sections of the line bundles ${\cal O}_X(-1,1,1,1)$ and ${\cal O}_X(2,-1,4,4)$, respectively. Here we are using the standard notation where the integers represent the coefficients in an expansion of the first Chern class of the line bundle in a basis provided by restricting the ambient K\"ahler forms to the Calabi-Yau manifold. As a somewhat larger example, in Section \ref{Sec:X3-3} we provide a case with Mordell-Weil rank 4.

As a final comment, it is clear that certain features -- such as the exact nature of the resolved geometry corresponding to an F-theory model, and thus its M-theory limit -- are obvious in this construction. In the more conventional approach to building F-theory models this information can be highly non-trivial to obtain and a lot of interesting work has been carried out in this regard (see Refs.~\cite{Esole:2011sm,Hayashi:2014kca,Esole:2014bka,Esole:2014hya,Braun:2014kla,Esole:2015sta,Braun:2015hkv,Esole:2016npg} for some recent advances).



\vspace{0.2cm}

The outline of this paper is as follows. In Section \ref{fibandsec}, we describe how to identify genus-one fibrations of CICYs and obtain explicit expressions for their sections, if present. In Section \ref{weierstrass}, we describe how to use this information to obtain the Weierstrass models and the Jacobians associated to the initial Calabi-Yau manifolds. We also give some more information on the resolved geometries and discuss the decomposition of the Picard lattice according to the theorem of Shioda-Tate-Wazir \cite{tsw}. Sections \ref{eg1}-\ref{Sec:X3-3} contain explicit examples of these techniques as applied to cases taken from the CICY data set. 

\section{Elliptic Fibrations and Sections} \label{fibandsec}

Consider a Calabi-Yau $n$-fold $X_n$ embedded as a complete intersection of $K$ hypersurfaces in a product of projective spaces $\mathcal A =  \prod_{r=1}^m \IP^{n_r}_{\bold x_r}$, where the subscripts $\bold x_r = (x_{r,0}: \cdots :x_{r,{n_r}})$ denote the homogeneous coordinates of the corresponding projective space $\IP^{n_r}$. Then, it can be characterized by a so-called configuration matrix,
\beq\label{Xn-conf}
X_n =\def\arraystretch{1.2}\left[\ba{c||ccc} 
\IP^{n_1}_{\bold x_1} &  a_1^1 & \cdots & a_K^1 \\
\IP^{n_2}_{\bold x_2} &  a_1^2 & \cdots & a_K^2 \\
\vdots & \vdots & \ddots &\vdots \\
\IP^{n_m}_{\bold x_m} &  a_1^m & \cdots & a_K^m 
\ea\right] \ .   
\eeq
Here the matrix entry $a_j^r$ denotes the degree of the $j$-th hypersurface equation in the homogeneous coordinates of the $r$-th projective space factor of the ambient space. The Calabi-Yau condition leads to the degree constraints,
\beq\label{deg-constraint}
n_r +1 = \sum\limits_{j=1}^K a_j^r \ , 
\eeq
for each $r=1, \cdots, m$, while the condition that the Calabi-Yau be an $n$-fold is given as
\beq
\sum\limits_{r=1}^m n_r = n+K \ . 
\eeq 
In this paper, we will be analyzing such Calabi-Yau manifolds realized as a CICY\footnote{In many computations we have made use of the ``CICY Package''~\cite{cicy}.}. Further, since we wish to study F-theory vacua, we will restrict our considerations to those CICYs with at least one ``obvious'' genus-one fibration, as we will describe in the next subsection. To avoid potential confusions, before we proceed, however, we clarify the terminology that we will use throughout the paper. ``Genus-one fibration'' refers to a fibration of genus-one curves, whether or not it has a section, while ``elliptic fibration'' implies the existence of a section. Note also that we will oftentimes hide the manifold subscripts that indicate dimensions, for example $X_n$ and $X$ will be used interchangeably unless confusions arise. 

\subsection{Genus-one Fibration Structures in CICYs} 
In general it is rather difficult to take a CICY configuration matrix and enumerate all of the ways in which the associated variety can be written as a genus-one fibration. However, there exists a class of genus-one fibrations which can be readily classified from the configuration matrix alone.

It is possible to perform arbitrary row and column permutations on a configuration matrix without changing the geometry that is described. These operations simply correspond to relabelling the $\mathbb{P}^{n_r}$ ambient factors and the hypersurface equations, respectively. Let us suppose that upon appropriate use of such permutations, the configuration matrix~\eqref{Xn-conf} can be put in the following block form, 
\beq\label{Xn-split-conf}
X_n =\def\arraystretch{1.2}\left[\ba{c||ccccc} 
\mathcal A_F &  O & \mathcal F \\
\mathcal A_B &  \mathcal B & \mathcal T \\
\ea\right]  \ ,  
\eeq
where $\mathcal A_F$ and $\mathcal A_B$ are two products of projective spaces such that 
\beq
\mathcal A_F \times \mathcal A_B  = \prod_{r=1}^m \IP_{\bold x_r}^{n_r} \equiv \cA  \ , 
\eeq
while $\mathcal F$, $\mathcal B$, $\mathcal T$, and $O =0$ are submatrices of the full configuration matrix. Such a configuration describes a fibration of the fiber $F = \left[ \mathcal A_F || \mathcal F \right]$ over the base $B=\left[\mathcal A_B || \mathcal B\right]$, where $\mathcal T$ describes the variation of the fiber over the base. Since the top-left block is a zero matrix, the Calabi-Yau condition~\eqref{deg-constraint} guarantees that the fiber $F$ obeys the analogous degree constraints and hence is Calabi-Yau itself too. Therefore, as long as the number of columns of $\mathcal F$ and the dimension of $\mathcal A_F$ are such that $F$ is of complex dimension $1$, the fibers will be Calabi-Yau one-folds: that is genus-one curves as desired. It follows that the base will then be of complex dimension $n-1$. 

Such genus-one fibration structures can easily be found at the level of configuration matrix via permutations of rows and columns, and will hence be referred to as ``obvious genus-one fibrations" (OGFs)\footnote{Note that in Ref.~\cite{Gray:2014fla} such OGFs were referred to as obvious elliptic fibrations (OEFs). In this paper, however, we need to carefully distinguish between fibrations with and without sections and thus will avoid this earlier nomenclature.}. Almost all CICY configurations have an OGF, and in many cases multiple such structures~\cite{Gray:2014fla, usscanning}. For example, among the $7,890$ CICY threefolds, $7,837$ can be brought into the OGF form~\eqref{Xn-split-conf} and a CICY threefold admits $9.85$ OGFs on average, with the number of OGFs per configuration ranging from $0$ to $36$. Similarly, among the $921,497$ CICY fourfolds, all but $477$ have an OGF and a CICY fourfold admits $54.6$ OGFs on average, with the number of OGFs per configuration ranging from $0$ to $354$. 

As an illustration of OGF, let us consider the $K3$ surface with the following configuration matrix,
\beq\label{K3-config-0}
\quad X_2 =\def\arraystretch{1.2}\left[\ba{c||cc} 
\IP^1_{\bold x_1} &  1 & 1 \\
\IP^2_{\bold x_2} &  1 & 2 \\
\IP^1_{\bold x_3} &  1 & 1 \\
\ea\right]  \ .
\eeq
The $K3$ surface $X_2$ admits an OGF structure over the base $B_1=\IP^1_{\bold x_3}$, where the configuration of the fiber is given as
\beq\label{Fiber-config-0}
\quad F =\def\arraystretch{1.2}\left[\ba{c||cc} 
\IP^1_{\bold x_1} &  1 & 1 \\
\IP^2_{\bold x_2} &  1 & 2 \\
\ea\right]  \ .  
\eeq
In this particular example the matrix ${\cal B}$ has $0$ columns. For the rest of this section and the next two, we will use this configuration as a simple and explicit example with which to clarify various techniques. The entire analysis for this $K3$ surface is put together in Section~\ref{eg1} in a self-contained manner for the reader who prefers a complete worked example to an illustrated general analysis. 

\vspace{0.1cm}

One further point that should be made before we proceed is that the OGFs just described need not be flat fibrations. As a simple example consider the following configuration, 
\beq\label{rps-Efib-21}
X_3 =\def\arraystretch{1.2}\left[\ba{c||cccccc} 
\IP^1_{\bold x_1} &  1 & 1 & 0  \\
\IP^1_{\bold x_2} &  1 & 0 & 1 \\ 
\IP^2_{\bold x_3} &  1 & 0 & 2  \\ 
\IP^2_{\bold x_4} & 1 & 1 & 1 \\
\ea\right]  \ .
\eeq
Here we can take the fiber to be described by
\begin{eqnarray} \label{mrthefib}
F=\def\arraystretch{1.2}\left[\ba{c||cccccc} 
\IP^1_{\bold x_1} &  1 & 1 & 0  \\
\IP^1_{\bold x_2} &  1 & 0 & 1 \\ 
\IP^2_{\bold x_3} &  1 & 0 & 2\ea\right] \;,
\end{eqnarray}
with $\IP^2_{\bold x_4}$ being the base.
Consider the defining relation given by the second column in Eq.~\eqref{rps-Efib-21}. This takes the following form,
\begin{eqnarray}
P_2=x_{1,0} l_0 ({\bf x}_4) + x_{1,1} l_1({\bf x}_4)\;,
\end{eqnarray}
where $l_0$ and $l_1$ are linear functions in the homogeneous coordinates of the base $\mathbb{P}^2_{\bold x_4}$. For general choices of complex structure, at a certain point in the base we have $l_0=l_1=0$. At such point $P_2=0$ holds automatically, and this equation does not provide a constraint in ${\cal A}_F=\mathbb{P}^1_{{\bf x}_1}\times \mathbb{P}^1_{{\bf x}_2}\times \mathbb{P}^2_{{\bf x}_3}$. Thus over that point in the base the fiber is actually two dimensional, not a curve, and thus the fibration is not flat. 

More generally, if there is any choice of point on the base, such that the associated choice of complex structure describing the fiber  Eq.~\eqref{mrthefib} is not a complete intersection, then the fibration will not be flat. For most of this paper, we will restrict ourselves to examining flat fibrations for ease and physical motivations. However, in Appendix \ref{rps}, we will apply our methods to a non-flat fibration and make some comments about the connection between non-flatness of fibrations and singularities in the associated Weierstrass model in which $f$ and $g$ vanish to orders $4$ and $6$.

Finally, it should be noted that although the statistics described above focused on OGFs in CICYs, an elliptic fibration structure can also be found via a set of criteria purely in terms of the intersection theory: 

\vspace{0.3cm}

\noindent \emph{Conjecture \cite{Kollar:2012pv}: Let $X$ be a Calabi-Yau $n$-fold. Then $X$ is genus-one fibered iff there exists a $(1,1)$-class $D$ in $H^2(X, \mathbb{Q})$ such that $(D \cdot C) \geq 0$ for every algebraic curve $C \subset X$, $(D^{{\rm dim}(X)})=0$ and $D^{{\rm dim}(X)-1} \neq 0$.}

\vspace{0.3cm}
\noindent For $n=3$ (i.e. a CY $3$-fold) this conjecture has been proven subject to the additional constraints that $D$ is effective or $(D \cdot c_2(X)) \neq 0$ \cite{oguiso,wilson}. It is straightforward to see in the case of many CICYs, including for example the $3$-fold given in \eref{intro1}, the criteria above can be used to verify that the OGF fibrations are in fact \emph{all fibrations} for the given CICY (see \cite{Hubsch:1992nu,Anderson:2008uw} for general formulas on intersection numbers and topology of CICYs). In summary, the CICY dataset provide a rich data set of examples of multiply genus-one fibered Calabi-Yau manifolds which we will exploit in Ref.~\cite{phys} with the techniques developed in this paper.

\subsection{Putative Sections to Genus-one Fibrations} \label{putsec}
Given a smooth, genus-one fibered Calabi-Yau $n$-fold, $X_{n} \to B_{n-1}$, the fibration may or may not have a section. Although many of the techniques in this paper apply also to the cases without a section, for a clearer interpretation of the corresponding F-theory vacuum, we will always start our geometric exploration by determining whether or not a section exists (see, e.g., Refs.~\cite{Braun:2014oya,Morrison:2014era,Anderson:2014yva,Klevers:2014bqa,Garcia-Etxebarria:2014qua,Mayrhofer:2014haa,Cvetic:2015moa,Grimm:2015ona,Martucci:2015dxa,Lin:2015qsa,Grimm:2015wda,Oehlmann:2016wsb,Cvetic:2016ner} for examples of recent work on the physics of genus-one fibrations without a section). We will separate the procedure of finding a section into two steps. In the first step we will find a candidate divisor class in which a section could live by imposing topological constraints. In the second step the section itself will be constructed explicitly as a map from the base to the fiber.

We call a codimension-one subvariety $S \subset X_n$ a ``putative section'' if it is an element of a divisor class $[S]$ that meets the following two criteria, both of which necessarily hold for a section: 

\begin{itemize}
\item [(a)] Let $\hat D^{\rm b}_{\alpha}$ for $\alpha=1, \cdots h^{1,1}(B)$ be a basis of divisor classes in the base and $D^{\rm b}_{\alpha}$ their pull backs to $X$. Then, due to Oguiso~\cite{oguiso}, the following equality between two intersection products, one on $X_n$ and the other on $B_{n-1}$, should hold,
\beq\label{intersection-c}
\left[S\right] \cdot  \prod\limits_{k=1}^{n-1}D^{\rm b}_{\alpha_k}  =  \prod\limits_{k=1}^{n-1} \hat D^{\rm b}_{\alpha_k} \ , 
\eeq
for each $(n-1)$-tuple $(\alpha_1, \cdots, \alpha_{n-1})$ with $\alpha_k = 1, \cdots, h^{1,1}(B)$. 
This is necessary if the fiber $F_p$ at a generic point $p \in B$ is to intersect $S$ at a single point. We will oftentimes refer to the criterion (a) as the Oguiso criterion.  

\item [(b)] For $S$ to be a section it should be birational to the base $B$, and thus necessarily obeys the following intersection criterion~\cite{Morrison:2012ei}, 
\beq\label{btop}
\left[S\right]\cdot \left[S\right] \cdot \prod\limits_{k=1}^{n-2}D^{\rm b}_{\alpha_k} = - \left[c_1(B)\right]\cdot \left[S\right] \cdot  \prod\limits_{k=1}^{n-2}D^{\rm b}_{\alpha_k}\ , 
\eeq
for each $(n-2)$-tuple $(\alpha_1, \cdots, \alpha_{n-2})$ with $\alpha_k = 1, \cdots, h^{1,1}(B)$. Note that this criterion has been well-established for $n=2$ and $3$, and is believed to also hold for CY $n$-folds with $n>3$. 
\end{itemize}
When applied to any particular case, the two criteria above lead to a family of possible divisor classes for putative sections. Given that any putative section $S$ is an element of a divisor class $[S]$, there exists a corresponding line bundle $\cO_X(S)$. In what follows we will frequently find it useful to consider these line bundles rather than the associated divisor classes.

\vspace{0.2cm}

As an illustration, let us consider the $K3$ configuration of Eq.~\eqref{K3-config-0}. If the first Chern class of $\cO_X(S)$ is $c_1({\cal O}_X(S)) =b_1 J_1 + b_2 J_2 + b_3 J_3$, where $J_{r=1,2,3}$ are the harmonic $(1,1)$-forms descending from the ambient projective pieces $\IP^{n_r}_{\bold x_r}$, then we will denote the line bundle as $\cO_X(S)=\cO_X(b_1, b_2, b_3)$. Here we have labeled the basis of forms such that the base $B=\IP^1_{\bold x_3}$ has a unique harmonic $(1,1)$-form generator $J_3$. The right hand side of criterion (a) is then simple to compute:
\beq \label{preeq0}
 \prod\limits_{k=1}^{1} \hat D^{\rm b}_{1} = \int_{\IP^1_{\bold x_3}} J_3= 1 \;.
\eeq
Here we have computed the intersection number of the divisor class in the base by integrating the dual $(1,1)$ form, which we take to be normalized in the usual manner, over that manifold. The left hand side of Eq.~\eqref{intersection-c} can be computed in terms of a similar integral over wedge products of dual forms -- this time over the Calabi-Yau manifold itself. Remembering that our unique divisor in the base is dual to $J_3$ and  how we are parameterizing the first Chern class of ${\cal O}_X(S)$ we find
\beq \label{preeq1}
\left[S\right] \cdot  \prod\limits_{k=1}^{1}D^{\rm b}_{1} = \int_X (b_1 J_1 + b_2 J_2 + b_3 J_3)  \wedge J_3 \ . 
\eeq
We will denote the form dual to the Calabi-Yau manifold itself, inside the ambient space, as $\mu_X$. In general this form can be read directly off from the configuration matrix \eqref{Xn-conf} as $\mu_X= \bigwedge_j a^r_{j} J_r$. Using this, we find that the integral in Eq.~\eqref{preeq1} can now be rewritten as follows, 
\begin{eqnarray} \label{preeq2}
\left[S\right] \cdot  \prod\limits_{k=1}^{1}D^{\rm b}_{1} &=&\int_{\IP^1_{\bold x_1} \times \IP^2_{\bold x_2} \times \IP^1_{\bold x_3}} \left[(b_1 J_1 + b_2 J_2 + b_3 J_3) \wedge J_3\right] \wedge \mu_X  \\ \nonumber &=&\int_{\IP^1_{\bold x_1} \times \IP^2_{\bold x_2} \times \IP^1_{\bold x_3}} \left[(b_1 J_1 + b_2 J_2 + b_3 J_3) \wedge J_3\right] \wedge \left[(J_1 + J_2 + J_3) \wedge (J_1 + 2J_2 + J_3) \right] \\ \nonumber
&=& 2b_1 +3 b_2 \ . 
\end{eqnarray}
Combining Eqs.~\eqref{preeq0} and \eqref{preeq2} we finally find that condition \eqref{intersection-c} results in the following constraint on the divisor class of any potential section, 
\beq\label{k3-cond1-0}
2 b_1 + 3b_2 \overset{!}{=} 1 \ .
\eeq

The criterion (b) can be written in terms of $b_r$ via a very similar intersection computation and leads to  
\beq\label{k3-cond2-0}
6 b_{1} b_{2} + 2 b_{2}^2 + 4 b_{1} b_{3} + 6 b_{2}b_{3} \overset{!}{=} -2 \ .
\eeq
These two putative-section criteria,~\eqref{k3-cond1-0} and~\eqref{k3-cond2-0}, can be solved as
\beq\label{gs-0}
b_{1} = -1-3k \ , \quad b_{2} = 1+2k \ , \quad b_{3} = 1+11k+14k^2 \ ,
\eeq
with an integer parameter $k \in \IZ$. For some small values of $k= \{1, 0, -1, -2\}$, for instance, we obtain the putative section classes, $\cO_X(-4,3,26), ~\cO_X(-1,1,1), ~\cO_X(2,-1,4), ~\text{and~}\cO_X(5,-3,35)$, respectively.

\subsection{Sections to Elliptic Fibrations}\label{section-to-efib}

To make further progress towards finding a section, beyond the necessary topological conditions imposed in the previous subsection, we must fix a complex structure for the Calabi-Yau manifold $X_n$. This is simply achieved by choosing an explicit set of defining equations $P_j({\bf x}_r)=0$ where $j=1,\ldots, K$, in terms of the homogeneous coordinates, $\bold x_r = (x_{r,0}:\cdots:x_{r,n_r})$, of the ambient space, $\cA = \prod_{r=1}^m \IP_{\bold x_r}^{n_r}$. 

Given a parameterization of putative section classes, such as Eq.~\eqref{gs-0}, and a choice of complex structure, the next step is to select a divisor $S_0$ satisfying criteria (a) and (b) which also has $h^0(X, \cO_X(S_0))=1$\footnote{It should be noted that the criterion (a) together with the line-bundle cohomology condition, $h^0(X, \cO_X(S))=1$, are not sufficient conditions for the existence of a section. In particular, it happens for some divisors that these two conditions are satisfied while the putative-section criterion (b) in Eq.~\eqref{btop} is not.}. The unique global holomorphic section (GHS), 
\beq\label{w-z}
z=z(\bold x_1, \cdots, \bold x_m) \in \Gamma (X,\; \cO_X(S_0))\ , 
\eeq
will then be explicitly found and proven to yield a true section to the fibration, whether holomorphic or rational.\footnote{Note that for the rest of the paper we will use the acronym GHS for global holomorphic section of a line bundle and reserve the word ``section'' for the (putative) section to a genus-one fibration.} 

The condition $h^0(X, \cO_X(S_0))=1$ follows, in the case where the section describes a smooth element of an effective divisor class, from the birationality of the desired section to the base. For an elliptically fibered Calabi-Yau manifold the cohomology of the trivial bundle over the base is always zero, except for the presence of a single global section. This is a simple consequence of the Calabi-Yau condition and the fact that pulling back any further cohomology elements under the projection map would lead to harmonic forms on the total space that are known not to exist \cite{Witten:1996bn}. Since these cohomologies are a birational invariant, and the section is birational to the base, we require that these statements should also hold for $S_0$. For a smooth section we can then use the Koszul sequence to relate trivial bundle cohomology on $S_0$ to that of ${\cal O}_X(S_0)$ as follows. The short exact Koszul sequence describing the section inside the Calabi-Yau gives rise to
\begin{eqnarray}
0\to  {\cal O}_{X}(-S_0) \to {\cal O}_X \to {\cal O}_X|_{S_0} \to 0 \;.
\end{eqnarray}
The associated long exact sequence in cohomology then contains the following piece.
\begin{eqnarray}
0\to H^3(X,{\cal O}_X(-S_0)) \to H^3(X,{\cal O}_X) \to 0
\end{eqnarray}
Using Serre duality on the Calabi-Yau manifold, together with $h^3(X,{\cal O}_X)=1$, we then find that $h^0(X, {\cal O}_X(S_0)) \stackrel{!}{=} 1$ as claimed\footnote{The condition $h^0(X,\cO_X(S_0))=1$ will in fact also hold in the case of singular sections. In the case of smooth sections, the higher cohomology groups also vanish: $h^{j}(X, \cO_X(S_0))=0$ $\forall j>0$. These vanishings are not a necessity for us however, since rational sections to an elliptic fibration may be singular, in which case $\cO_X(S_0)$ may have a non-trivial higher cohomology (see Eq.~\eqref{ps2} for an example). We do not add the condition $h^0(X,{\cal O}_X(S_0))=1$ to our definition of a putative section, since the cohomology condition does not lead to a closed-form constraint unlike the criteria (a) and (b) and thus it is practically employed in a different manner.}.

It should be noted that putative-section criteria (a) and (b), even when equipped with the cohomology condition, $h^0(X, \cO_X(S_0))=1$, do not guarantee that the zero locus of Eq.~\eqref{w-z} is a section to the genus-one fibration. Therefore, in order to ensure that the putative section $S_0$ leads to a true section, it is desired to provide an explicit rational map from $B$ to $S_0$, specified by parametrization of the form, 
\beq\label{rat}
\bold x_F = \bold x_F(\bold x_B) \ , 
\eeq
where $\bold x_F$ and $\bold x_B$ collectively denote the ambient homogeneous coordinates of $\mathcal A_F$ and $\mathcal A_B$, respectively. Once an explicit expression for the GHS is found in terms of the ambient coordinates $\bold x_r$, as implied by Eq.~\eqref{w-z}, one can indeed solve for the fiber coordinates in terms of the base coordinates (up to some subtleties which we will discuss further in the Example Sections). However, the Oguiso criterion (a), together with the fact that the intersection numbers in the projective-space product $\mathcal A_F$ are non-negative, tend to force the line bundle $\cO_X(S_0)$ to simultaneously have positive and negative degrees along the $\mathcal A_F$ direction. For example, in the $K3$ case above, the resulting constraint was Eq.~\eqref{k3-cond1-0}, which indeed forces one of $b_1$ and $b_2$ to be negative and the other positive. Thus, the GHS of $\cO_X(S_0)$ cannot be written as a polynomial in $\bold x_r$. Instead, it takes a rational form and can be constructed using the techniques developed in the context of generalized CICYs~\cite{Anderson:2015iia}.

Let us briefly review how such a rational expression for the GHS of $\cO_X(S_0)$ can be obtained, in the presence of both positive and negative degrees. We first label the line bundle $\cO_X(S_0)$ in terms of its first Chern class as $\cO_X(b_1, \cdots , b_m)$ as in Section \ref{putsec}. We then take the following rational ansatz for its GHS, 
\beq\label{ansatz}
\frac{N(\bold x_1, \cdots, \bold x_m)}{D(\bold x_1, \cdots, \bold x_m)} \in \Gamma(X, \cO_X(b_1, \cdots, b_m))\ , 
\eeq
where $N$ and $D$ are polynomials in $\bold x_1, \cdots, \bold x_m$ of multi-degree $(b_1, \cdots, b_m)_+$ and $(b_1, \cdots, b_m)_-$, respectively. Here, the signs in subscript indicate that only the degrees with the specified sign are extracted (and are flipped to be positive for the `$-$' case). For example, for $\cO_X(-1,1,1)$, which is one of the putative sections we found for our simple K3 case (\ref{K3-config-0}-\ref{Fiber-config-0}), the polynomials $N$ and $D$ are of multi-degrees $(-1,1,1)_+=(0,1,1)$ and $(-1,1,1)_-=(1,0,0)$, respectively. 
For a given denominator $D$ with the right degree, one is not allowed to choose a generic numerator $N$. This is because the ratio $N/D$ would then behave irregularly at a generic point where $D$ vanishes, while the GHS of the line bundle ${\cal O}(-1,1,1)$ is known to be associated to a polynomial in coordinates on $X$ itself which can not exhibit any such singularities. Instead $N$ must be chosen such that on the complete intersection $X$ the vanishing of $D$ is completely cancelled by that of $N$, yielding a regular function. 
Such a tuning of coefficients in $N$ provides an explicit GHS in terms of $\bold x_r$. 

\vspace{0.2cm}

As an illustration, let us go back to the $K3$ example and take the putative section $\cO_X(-1,1,1)$ from Eq.~\eqref{gs-0} with $k=0$, whose cohomology is computed as $h^\bullet (X, \cO_X(-1,1,1)) = (1,0,0)$. To find the (unique) rational expression of the form~\eqref{ansatz}, where $D(\bold x_1)$ is linear in $\bold x_1$ and $N(\bold x_2, \bold x_3)$ is bi-linear in $\bold x_2$ nd $\bold x_3$, we need to make use of the first defining equation of $K3$, which we denote by $P_1(\bold x_1, \bold x_2, \bold x_3)$. This defining equation can be expanded as 
\beq\label{k3-P-0}
P_1=x_{1,0} \, p_1(\bold x_2, \bold x_3) + x_{1,1} \, p_2(\bold x_2, \bold x_3) \ , 
\eeq
for bi-linear polynomials $p_1$ and $p_2$ (note from the first column of Eq.~\eqref{K3-config-0} that $P_1$ is tri-linear). 
Without loss of generality, we may choose to use $D(\bold x_1)=x_{1,0}+10x_{1,1}$, for example, and proceed to determine the bi-linear polynomial $N(\bold x_2, \bold x_3)$, that vanishes whenever $D$ does in $X$. Substituting the solution $x_{1,0} = - 10 x_{1,1}$ to $D(\bold x_1) =0$ into $P_1$ we immediately have a perfect candidate for a numerator, $-10 p_1 + p_2$ which vanishes on $X$ whenever $D$ does. This particular choice of $N$ can be thought of as a tuning of the coefficients of the six bi-linear monomials in $\IP^2_{\bold x_2} \times \IP^1_{\bold x_3}$. Thus, the GHS in this case is constructed (uniquely as a function on $X$) as
\beq\label{equivalent}
\frac{N(\bold x_2, \bold x_3)}{D(\bold x_1)}=\frac{-10 p_1(\bold x_2, \bold x_3) + p_2 (\bold x_2, \bold x_3)}{x_{1,0} + 10 x_{1,1}} \ .
\eeq
In particular, this expression can be proven equivalent when evaluated on $X$ to any other expression obtained from a different choice of the denominator polynomial $D$ \cite{Anderson:2015iia}. In some cases, the GHS can have sufficiently complex dependence on the defining equations of the CY that it can be difficult to determine the section analytically along the lines above.  In this case it is still possible to determine the appropriate regular rational function numerically. This can be done by locating a large number of points on the denominator, $D$ -- found by intersecting the CY defining equations, together with the denominator, with an appropriate number of generic multi-linear hypersurface constraints. By requiring that the numerator also vanish (along with the denominator) for this collection of points, the coefficients of a generic numerator can be fully fixed, leading to a complete description of the GHS \cite{Anderson:2015iia}.  

\vspace{0.2cm}

It is worth mentioning that there is a related method for the numerator tuning, which shares the same spirit as the previous method, and which is applicable specifically when the base $X$ of the line bundle in question is a fibration itself (as in our case). Again, one starts from the ansatz~\eqref{ansatz}, together with a choice of $D$ with the right degree. At a generic point $p \in B$, $D=0$ can be solved for a discrete set of points on the fiber $F_p$ over $p$. Then, the numerator polynomial $N$, when evaluated at each of these fibral points, should vanish. By choosing sufficiently many points on the base $B$ one can then obtain enough constraints on the coefficients in $N$ to uniquely determine the numerator. Despite being essentially the same as the first method described above, this approach benefits from the fact that one does not need to spot the correct combination of defining relations that must be used in order to derive the appropriate numerator. This method therefore lends itself better to automation on a computer. This alternative method will be used in Section~\ref{eg1} for the same $K3$ geometry and will be shown to give the same global section expression as in Eq.~\eqref{equivalent}.

Using any of the above methods, one can obtain the GHS expression and hence, also the explicit section map of the form~\eqref{rat}. If the section map can explicitly be shown to be rational this way, we will have found a legitimate section. It should be noted, however, that the corresponding divisor is not necessarily smooth. For example, putative sections may be reducible, containing a genuine section as an irreducible component as well as a vertical divisor therein (see Appendix~\ref{rps} for explicit examples); these cases can (and will) be ruled out by testing the irreducibility of the divisor. In general, even a genuine section can be a singular divisor and some CY geometries presented in this paper will include such instances. 

As has been mentioned above, the sections that are found may be rational and thus wrap fibral $\IP^1$'s over certain points of the base. For the case of smooth sections, therefore, where the Euler number $\chi(S_0)$ of $S_0$ is well-defined, this number should be closely related to that of the base $\chi(B)$ via the manner in which those $\IP^1$'s are wrapped. In particular, for $X_2 \to B_1$, the sections are always holomorphic, while for $X_3 \to B_2$, the sections may wrap $\IP^1$'s over a finite number of base points. For the latter case, the Euler number difference, $\chi(S_0)-\chi(B)$, counts the total number of fibral $\IP^1$'s being wrapped by the rational section, $S_0$. All of these statements will be confirmed for explicit examples in later Sections. 

Finally, let us fix some nomenclature. Since Eq.~\eqref{w-z} is to define the $z$ coordinate of the Weierstrass model in Section~\ref{w-procedure}, we denote the associated line bundle as
\beq\label{lz}
L_z := \cO_X(S_0) \ .
\eeq
Likewise, we further define the two line bundles,
\bea\label{lx}
L_x &:=& \cO_X (2S_0)\otimes K_B^{-2} \ , \\ \label{ly}
L_y &:=&\cO_X(3S_0)\otimes K_B^{-3} \ , 
\eea
where $K_B$ is the canonical bundle of the base $B$ pulled back to $X$. The remaining Weierstrass coordinates $x$ and $y$ will then be constructed as GHS's of $L_x$ and $L_y$, respectively, in manner similar to the method described above. This will be discussed further in Section~\ref{w-procedure}. 


\section{Singular Fiber Analysis} \label{weierstrass}
In order to understand the F-theory effective physics associated to an elliptically fibered Calabi-Yau manifold, $X$ it is necessary to obtain a minimal limit of the geometry in which the fibers are irreducible and possibly singular. The structure of the singular fibers encodes information about non-Abelian gauge symmetries, charged matter and more in the effective theory \cite{Vafa:1996xn,Morrison:1996na,Morrison:1996pp}.

In general, for an elliptically fibered CY $n$-fold, $X \to B$, with section there are three possible routes to a ``minimal" form the geometry suitable for an F-theory limit. 1) By Nakayama's theorem, any such elliptic fibration is birational to a Weierstrass model \cite{nakayama}. 2) To $X$ we can associate the Jacobian, $J(X)$ \cite{art} and finally 3) All reducible components of fibers can be explicitly blown down to form a ``minimal model" in the sense of the Minimal Model Program (MMP) (see e.g. \cite{hacon}). In the case of smooth, elliptically fibered CY $2$-folds these three procedures all lead to the same simple/minimal geometry. However, in the case of CY $n$-folds with $n \geq 3$ these approaches can differ (and ``minimal models" in the sense of the MMP are non-unique). For example, while the Weierstrass models of smooth CY $3$-folds are birational to $X$, the topology of $J(X)$ and $X$ may differ still further. Moreover, all three approaches can lead to different singular fibers at codimension 2 and higher in the base. As a result, our focus will be primarily on Weierstrass models as constructed by Nakayama \cite{nakayama}, however, as we will discuss below, the discriminant loci, $\Delta \subset B$, of these different forms are in fact identical and this observation, as well as information from the different approaches will be used to simply extract the gauge symmetries and charged matter from the original geometry.

We summarize below the ways that these three different approaches can be used to analyze the singular fibers of $X$. The first approach will be to construct a Weierstrass model $X_W$ (built as a hypersurface defined in an ambient $\mathbb{P}^2$-bundle defined by the projectivization of three line bundles over $B$), the second will be to form the Jacobian $J(X)$ of $X$ (or that of its blow down), and the third will be to analyze the smooth (resolved) geometry $X$  itself in order to directly study the singular fibers. 
A key object in studying the singular fibers of our manifolds is the discriminant locus for the elliptic fibration, and the following is expected of the triple of geometries, $X_W$, $J(X)$, and $X$: 
\bi
\item $X$ is birational to its Weierstrass model $X_W$. Since the singular fibers of $X$ remain singular after blowing down any $\IP^1$'s therein, the discriminant loci of the two elliptic fibrations agree as an algebraic variety in the base $B$. We then have, in particular,
\beq
\Delta_{\rm res} = \Delta_W \ ,
\eeq
where $\Delta_{\rm res}$ and $\Delta_W$ denote the discriminant polynomials in $B$ for $X$ and $X_W$, respectively.
\item The Jacobian\footnote{To avoid confusion it should be noted that in the mathematics literature, the ``Jacobian" of a genus one fibered manifold $X \to B$ is sometimes taken to refer to \emph{any} fibration of the form $X \to B$ with a section. Here we will reserve the terminology of ``Jacobian" (or $J(X)$) to refer explicitly to a fibration constructed via the variable changes/procedure outlined in \cite{art,Braun:2011ux, Braun:2014qka}.} $J(X)$ of $X$ is constructed as a fibration over the same base $B$, in such a way that the fiber over each point is the Jacobian of the curve defined by the original fiber (that is, the moduli space of degree zero line bundles on that curve). This construction insures that the $j$-invariant of each fiber of $J(X)$ agrees with that of the original fiber of $X$. The two discriminant loci in $B$ are identified and it is expected that 
\beq
\Delta_{\rm res} = \Delta_J \ , 
\eeq
where $\Delta_J$ denotes the discriminant polynomial in $B$ for $J(X)$. Furthermore, at generic points on a codimension-one locus in the base, the singular fibers have the same Kodaira type \cite{kodaira} in both geometries. 
As described above, the behaviors of the singular fibers at codimension two (or higher) may differ in general. 
\item Due to the two points made above, the Weierstrass model $X_W$ and the Jacobian $J(X)$ also share the discriminant loci in $B$ and we have
\beq
\Delta_W = \Delta_J \ .
\eeq
Furthermore, the singular fibers of $X_W$ and $J(X)$ are of the same Kodaira type generically at a codimension-one locus in the base.
\ei

In what follows, we will sketch how the codimension-one locus in the base is obtained for each of the three approaches in turn, oftentimes returning to our simple $K3$ example whenever illustration is in need.

\subsection{Weierstrass Models}\label{w-procedure}
An elliptic fibration with a section can be associated to a Weierstrass model. Here we will follow a procedure due to Deligne (for elliptic curves) and Nakayama (for elliptic fibrations) \cite{deligne,nakayama} which is well known in the physics literature (see Refs.~\cite{Ovrut:2000qi,Morrison:2012ei} for some explicit examples of its application in such contexts)\footnote{We would like to thank T. Pantev for very useful conversations about algorithmically applying this procedure.}. Schematically, a Weierstrass model for an elliptic fibration is built as a hypersurface constraint (with cubic fiber) within a projectivization of three line bundles $\mathbb{P}(\cO_{B} \oplus {\cal L}^{2} \oplus {\cal L}^3)$. The Calabi-Yau condition on $X$ fixes ${\cal L}=K_{B}^{-1}$ and a change of variables makes it possible to describe the elliptic fiber as a degree $6$ hypersurface in a weighted $\mathbb{P}_{123}$. In this description then, the fiber coordinates $z,x,y$ (of weights $(1,2,3)$) are associated to global holomorphic sections of the following line bundles over $X$:
\beq
z\sim \cO(S_0)~~,~~x \sim \cO(2S_0) \otimes {\cal L}^2~~,~~y\sim \cO(3S_0) \otimes \cL^3
\eeq
 
We first choose a zero section, with respect to which a Weierstrass model will be found, and denote the associated line bundle as $L_z$, which in particular satisfies $h^0(X, L_z)=1$.  
\begin{itemize}
\item  Using the technique reviewed in the previous Section (see the paragraph including Eq.~\eqref{ansatz}), we can obtain the rational expression for the unique GHS of $L_z$. This will be the $z$ coordinate of the Weierstrass form, expressed in terms of the coordinates of our original ambient space:
\begin{eqnarray}
z=z(\bold x_1, \cdots, \bold x_m) \;.
\end{eqnarray}
\item Next we take the line bundle $L_x$ as defined in Eq.~\eqref{lx}. The dimension of the space of global sections of this bundle, $h^0(X,L_x)$, is such that it is one larger than the subset of elements of that space that is spanned by a basis that can be written as $z^2$ multiplied by polynomials in the base coordinates. In essence, that additional element will describe the $x$ coordinate of the Weierstrass model in terms of the original ambient space coordinates.

In practice one may take a generic element of $H^0(X, L_x)$ to describe $x$. This is simply because such a generic element will indeed not be proportional to $z^2$, containing some portion of the remaining basis element in the cohomology group. There are of course a plethora of different generic elements that could then be chosen. This freedom simply corresponds to making different choices of the Weierstrass coordinate $x$, which are related under coordinate transformations mixing $x$ and $z$ which maintain the Weierstrass form. After choosing a generic element of $H^0(X, L_x)$ we can now employ the methodology described in the proceeding section to obtain an explicit description of the $x$ coordinate of the Weierstrass form.
\begin{eqnarray}
x&=&x(\bold x_1, \cdots, \bold x_m)
\end{eqnarray}

\item The final Weierstrass coordinate $y$ is obtained in a very similar manner as a generic element of $H^0(X, L_y)$. In this instance, the dimension of the cohomology group, $h^0(X, L_y)$, is such that it is one larger than the subset of elements of that space that is spanned by a basis that can be written as $z^3$ or $xz$ multiplied by a polynomial in the base. As before, different choices for the coordinate $y$ will lead to Tate forms for the fibration which are related by coordinate transformations mixing $x, y$ and $z$ which leave the Weierstrass form invariant. Once more, we employ the technology of the previous section to obtain the explicit description of the coordinate at hand in terms of those of the original description of the ambient space of the manifold.
\begin{eqnarray}
y&=&y(\bold x_1, \cdots, \bold x_m) 
\end{eqnarray}

\item Finally, given that we now have explicit expressions for $x$, $y$ and $z$ in terms of our original ambient space coordinates, we can now, by straightforward calculation, find a relationship between them that is of the Tate form \cite{tate} (up to scaling),
\beq\label{wcubic}
y^2 + c_1 x y z + c_3 y z^3 + c_0 x^3 + c_2 x^2 z^2 + c_4 x z^4 + c_6 z^6  = 0 \ .
\eeq
Here, the $c_i$'s are GHS's of $K_B^{-i}$, that is functions of the base coordinates of specific degrees. That there is a unique such relation follows from similar arguments to those given in the previous bullet points. The left hand side of the relation Eq.~\eqref{wcubic} is associated to an element of $H^0({\cal O}_X(6 S_0)\otimes K_B^{-6})$. The dimension of this space is one less than that naively spanned by elements of appropriate degree that can be written as $y^2$, $xyz$, $y z^3$, $x^3$, $x^2 z^2$, $x z^4$ and $z^6$ multiplied by elements of the relevant $H^0(X,K_B^{-i})$. Thus there must be one relation between these quantities which vanishes as in Eq.~\eqref{wcubic}.

Practically to find the relationship in Eq.~\eqref{wcubic} we employ a similar technique to that discussed in the paragraph under Eq.~\eqref{equivalent}. We write out a generic relation of the correct form with undetermined numerical coefficients. In particular, we form a basis $M^{l_i}$ of $H^0(X,K_B^{-i})$ where $l_i=1,\ldots, h^0(X,K_B^{-i})$. Expanding the $c_i$'s we then have the following,
\begin{eqnarray}
c_i = c_{i \; l_i} M^{l_i} \;,
\end{eqnarray}
which is taken to define the $c_{i \; l_i}$ -- our numerical coefficients.

Writing  Eq.~\eqref{wcubic} in terms of the $c_{i \; l_i}$ and the ambient coordinates of the original description of the manifold, we then substitute in coordinates of a point on the manifold (solved for using our original description of the space) to obtain a relationship between the $c_{i \; l_i}$. Repeating this procedure with sufficient numbers of points on the manifold we obtain a system of linear equations in the numerical coefficients which, by a naive counting of equations, would seem to be over-constrained. Nevertheless, this system can then be solved to uniquely determine the $c_{i \; l_i}$, and thus the relationship Eq.~\eqref{wcubic}. The fact that such parameter values can be found satisfying the equation system is, in itself, a reassuring check of the method.
\end{itemize}

\vspace{0.2cm}

As an illustration, let us return to the $K3$ geometry with configuration~\eqref{K3-config-0} with the line bundle $L_z = \cO_X(-1,1,1)$, which has already been proven to give a section to the elliptic fibration. The GHS $z$ of $L_z$ can be obtained explicitly as Eq.~\eqref{equivalent}, and similarly, one may easily find the generic GHS's associated to $x$ and $y$ using the same technique. Here we will simply note that, in performing this computation, since $K_B^{-1}={\cal O}(0,0,2)$ and ${\cal O}_X(S_0)=\cO_X(-1,1,1)$, the other relevant line bundles are given as $L_x= {\cal O}_X(2 S_0) \otimes K_B^{-2} ={\cal O}_X(-2,2,6)$ and $L_y= {\cal O}_X(3 S_0) \otimes K_B^{-3} ={\cal O}_X(-3,3,9)$, respectively.  Similarly, the relationship Eq.~\eqref{wcubic} is associated with an element of the zeroth cohomology of ${\cal O}_X(6 S_0) \otimes K_B^{-6} = {\cal O}_X(-6,6,18)$. We omit the explicit expressions for $x$ and $y$ here, and the detailed form of the final Weierstrass form due to their length. More details can be found for this specific example in Section \ref{K3-wm}.

\vspace{0.2cm}

Once the relation~\eqref{wcubic} is obtained, via an appropriate rescaling of $x$ we can put Eq.~\eqref{wcubic} into the standard Tate form, 
\beq\label{tate}
y^2+a_1 x y z+ a_3 y z^3= x^3 + a_2 x^2 z^2 + a_4 x z^4+ a_6 z^6\ , 
\eeq
where $a_i \in \Gamma(B, K_B^{-i})$, and then also into the Weierstrass form, 
\beq\label{weier}
y^2 = x^3 + f_W x z^4+ g_W z^6\ , 
\eeq
where 
\bea
f_W&=&-\frac{1}{48} (b_2^2 - 24 b_4) \ , \\ 
g_W&=&-\frac{1}{864}(-b_2^3 + 36 b_2 b_4 -216 b_6) \ , 
\eea
with 
\bea
b_2 &=& a_1^2 + 4 a_2 \ , \\
b_4 &=& 2 a_4 + a_1 a _3 \ , \\
b_6 &=& a_3^2 + 4 a_6 \ .
\eea
In particular, the discriminant polynomial for such a Weierstrass model is given as 
\beq
\Delta_W = 4 f_W^3 + 27 g_W^2  \ . 
\eeq

Once the Tate form/Weierstrass form has been obtained one can use the standard techniques in order to analyze the singular fibers.

\subsection{Jacobians} \label{mdelajac}

In the case where the fiber of a CICY is realized as a complete intersection of codimension one or two, the Jacobian of $X_n$ can be formed by using the results in Refs.~\cite{art, Braun:2011ux, Braun:2014qka}\footnote{The results of~\cite{Braun:2014qka} can be searched from the following website: http://wwwth.mpp.mpg.de/members/jkeitel/Weierstrass/}. This work provides a list of Jacobians for all elliptic fibers realized as a complete intersection of codimension $1$ or $2$ in any toric variety, thus including products of projective spaces as a special case. More generally, however, the fibers we will encounter can be of higher codimension. Rather than generalizing the results of Refs.~\cite{art, Braun:2011ux, Braun:2014qka} in these cases, we find it more expedient to blow down the fiber until it reaches a codimension one or two description by utilizing the process of ``contraction" \cite{Candelas:1987kf}.

Contraction refers to the procedure of making the configuration matrix smaller by removing a row of $1$'s as follows, 
\beq\label{contraction}
\def\arraystretch{1.2}\left[\ba{c||cccc} 
\IP^a &  1 & \cdots & 1 & 0 \\
\mathcal A' &  \bold u_1 & \cdots & \bold u_{a+1} & C \\
\ea\right] 
\quad \longrightarrow\quad 
\def\arraystretch{1.2}\left[\ba{c||cc} 
\mathcal A' &  \sum\limits_{i=1}^{a+1} \bold u_i & C \\
\ea\right]  \ . 
\eeq
Here the first $a+1$ polynomials have merged to a single determinantal polynomial. We perform this procedure in such a manner that, while the description of the fiber in Eq.~\eqref{Xn-split-conf} is changed, the description of the base, $B=\left[\mathcal A_B || \mathcal B\right]$, remains invariant.  As has been described in Ref.~\cite{Candelas:1987kf}, this procedure corresponds to blowing down $\IP^1$'s in a manner that may or may not be associated with a geometric transition. In the case where the final manifold is different from the initial one we say the contraction is effective and otherwise it is ineffective. In our case we are clearly blowing down $\IP^1$'s in the fiber.

Even though the final manifold after completing this contraction may be different from our starting configuration it will have the same discriminant locus in the same base (and thus, so will its Jacobian). This is simply because the $\IP^1$'s in the fiber that are being blown down are singular before and after the process and therefore project to a point on the discriminant in both cases. Since the discriminant is what we are trying to obtain here, we are able to contract to get a codimension one or two fiber and then make use of the aforementioned existing results without any loss of information. 

In making use of the results of Refs.~\cite{art, Braun:2011ux, Braun:2014qka}, for a given complex structure for $X_n$ (or its blow down), we consider the defining equations for the fiber as polynomials in the fibral homogeneous coordinates, demoting the base coordinates to parameters. Then the Jacobian of the form, 
\beq
y^2 = x^3 + f_J x z^4+ g_J z^6 \ , 
\eeq
can be immediately read off, and its discriminant locus is obtained in turn by the zero locus of
\beq
\Delta_J = 4 f_J^3 + 27 g_J^2 \ , 
\eeq
in the normal way. This is a polynomial in the base coordinates that we now promote to variables again. 

\vspace{0.2cm}

For an illustration, let us return to our toy example of the $K3$ configuration~\eqref{K3-config-0}. 
Its fiber configuration~\eqref{Fiber-config-0}, 
\beq
\quad F =\def\arraystretch{1.2}\left[\ba{c||cc} 
\IP^1_{\bold x_1} &  1 & 1 \\
\IP^2_{\bold x_2} &  1 & 2 \\
\ea\right]  \ ,
\eeq 
is already a complete intersection of codimension two in $\mathcal A_F = \IP_{\bold x_1}^1 \times \IP_{\bold x_2}^2$. The defining relations are bi-degree $(1,1)$ and $(1,2)$ polynomials and this codimension-two fiber has the PALP ID $(4,0)$, which, via the result of Ref.~\cite{Braun:2014qka}, can straightforwardly be transformed into the Jacobian. The two defining polynomials, due to the base twist in the full configuration~\eqref{K3-config-0}, also depend on $\bold x_3$ in our case. However, we demote these variables to parameters so that the various monomial coefficients in the defining equations can be thought of as a polynomial parameterized by the base coordinates $\bold x_3$. The expressions for $f_J$ and $g_J$ in terms of those monomial coefficients are immediately found and thereby one obtains $f_J \in \Gamma (B, K_B^{-4})$ and $g_J \in \Gamma(B, K_B^{-6})$ explicitly. 

In every case we have computed the discriminant obtained in this manner has matched that obtained by the other two methods discussed in this section. This is a highly non-trivial check of the above procedure, several explicit examples of which will be provided in later sections.

\subsection{Resolved Geometries}\label{res-geom}

One of the benefits to the approach being espoused here for constructing F-theory compactifications is that the resolved space associated to the models is known from the start. Bertini's theorem (see Refs.~\cite{ha, gh}) guarantees that a CICY of the form being considered is smooth, presuming of course that a generic enough complex structure is chosen. 

In the original description of the manifold, the discriminant locus of the fibration can be explored by directly computing where over the base the fiber becomes singular. Starting with the configuration matrix of the form \eqref{Xn-split-conf},   
\beq\label{fiber-conf1}
X_n =\def\arraystretch{1.2}\left[\ba{c||ccccc} 
\mathcal A_F &  O & \mathcal F \\
\mathcal A_B &  \mathcal B & \mathcal T \\
\ea\right] \ , 
\eeq 
one can perform such a computation as follows.

\begin{itemize}
\item We will denote the coordinates of the base ambient space, ${\cal A}_B$ by $\bold x_B$, and the coordinates of the fiber ambient space, ${\cal A}_F$ by $\bold x_F$.
\item We write as $P_{\hat{j}}$ those defining relations associated to the last block of columns in Eq.~\eqref{fiber-conf1}, that is those defining equations associated to the following portion of the configuration matrix.
\beq\label{fiber-conf}
X_n =\def\arraystretch{1.2}\left[\ba{c||ccccc} 
\mathcal A_F  & \mathcal F \\
\mathcal A_B  & \mathcal T \\
\ea\right] \ . 
\eeq 
\item We then form the equation system
\beq\label{fiber-sing}
P_1=\cdots = P_K = 0 \ , \quad \wedge_{\hat{j}} {\rm d}_F P_{\hat{j}} = 0  \ , 
\eeq
where the exterior derivative ${\rm d}_F$ is only taken with respect to the variables $\bold x_F$ and not with respect to $\bold x_B$. The differentiated conditions here describe when the normal form to the fiber is ill defined -- that is they describe for what values of the ambient space coordinates the fiber becomes singular. Including the defining relations in the equation system then gives us the points on the Calabi-Yau itself where the fiber becomes singular.
\item Finally we want to project the equation system Eq.~\eqref{fiber-sing} to obtain those points on the base above which their are singularities in the fiber. This projection is equivalent algebraically to the process of elimination. We must eliminate the variables $\bold x_F$ to obtain a necessary and sufficient set of relations on the $\bold x_B$ such that there is a solution to Eq.~\eqref{fiber-sing} for some value of the fiber coordinates. Thinking of the equations as generators of an ideal we wish to form
\begin{eqnarray}
\left<P_1,\ldots, P_K,\wedge_{\hat{j}} {\rm d}_F P_{\hat{j}} \right> \cap \mathbb{C}[{\bf x_B}]\;.
\end{eqnarray}
Here by an abuse of notation we have denoted the polynomial coefficients of an expansion of $\wedge_{\hat{j}} {\rm d}_F P_{\hat{j}}$ in a basis of forms of an appropriate degree by the expression itself. Such an elimination can easily be performed with a Gr\"obner basis computation and results in a set of equations in the variables $\bold x_B$ describing the discriminant locus in the base of the fibration.
\end{itemize}

The last step in this procedure, the elimination process, is computationally expensive in large examples. Nevertheless, the full description~\eqref{fiber-sing} of location of fiber singularities in the total space of the Calabi-Yau manifold is already useful in comparing to computations performed using the proceeding methods described in this section. We will return to comparing the discriminants we have found in various ways, which, as expected, have matched in every case we have investigated, in the Example Sections to follow. It should also be noted that, given a discriminant locus derived using one of the other methods in this section, one could use the initial description of the manifold to investigate the nature of the associated singular fibers. One would simply choose values of $\bold x_B$ which lie on the discriminant and substitute these into the $P_{j}$ to obtain an explicit description of the singular fiber as a variety in ${\cal A}_F$. We will return to this point later. 

\section{The Mordell-Weil Group}

\subsection{Decomposition of the Picard Lattice} \label{picsec}
For an elliptic Calabi-Yau manifold $X_n$ with a section, the Shioda-Tate-Wazir theorem~\cite{tsw} states that the Picard lattice of $X_n$ is generated by linearly independent basis elements of the following four types\footnote{The statement is for the Ner\'on-Severi lattice, which coincides with the Picard lattice for a Calabi-Yau manifold.}: (1) base divisor classes pulled back to $X_n$, (2) ``fibral" divisors associated to blow-ups in the fiber ({\it i.e.}, vertical divisor classes that are not pulled back from the base), (3) the zero section, and (4) a basis of the rational sections generating (the free part) of the Mordell-Weil (MW) group (the additive group of sections to the elliptic fibration -- see, for example, Ref.~\cite{ec}). 
This in particular implies the following dimensional relation for $n\geq 3$:
\beq\label{box}
h^{1,1}(X_n) = h^{1,1}(B_{n-1}) + \sum\limits_{A}{\rm rk}\, G_A + 1 + {\rm rk}\,MW(X_n) \ .
\eeq
Here $G_A$ are the non-abelian Lie groups, each associated with the reducible fiber type over an irreducible component of the discriminant locus in $B_{n-1}$, and $MW(X_n)$ denotes the Mordell-Weil group of $X_n$.  

It is worth emphasizing how useful Eq.~\eqref{box} is for our purposes. Given a specific configuration for $X_n$ and $B_{n-1}$ in the form~\eqref{Xn-split-conf}, it is a straightforward exercise in algebraic topology to compute $h^{1,1}(X_n)$ and $h^{1,1}(B_{n-1})$. 
Furthermore, factorization of the discriminant equation is straightforward, from which one can easily read off the enhancement pattern of the fiber singularity, and in particular, ${\rm rk}\,G_i$. 
On the other hand, determination of ${\rm rk}\, MW(X_n)$ involves a careful analysis of the section structure, which in many cases is a difficult task. Thus, in analyzing the MW group structure, the relation~\eqref{box} can be used as either a consistency check on a direct computation or an indirect method to determine the MW rank, as will be illustrated with examples in later sections. 

From the physical perspective, Eq.~\eqref{box} also plays an important role in systematic exploration of the F-theory vacua from the plethora of elliptically fibered CICY threefolds and fourfolds~\cite{usscanning, Gray:2014fla}. Upon compactifying F-theory over an elliptic Calabi-Yau manifold $X_n$, one obtains a $(12-2n)$-dimensional effective theory with gauge group of the form,
\beq
G= U(1)^{{\rm rk}\, MW(X)} \times \prod\limits_{A} G_A \ .
\eeq
Since $h^{1,1}(X_n)$ and $h^{1,1}(B_{n-1})$ can be computed in a systematic manner for CICYs, Eq.~\eqref{box} makes it easy to classify the F-theory vacua with a fixed total rank, ${\rm rk}\, G$, of the gauge group. Furthermore, a relatively straightforward analysis of the discriminant locus and of the enhancement pattern of the fiber singularities of the manifold can be used to determine the non-abelian part of the gauge group in a systematic manner. Thus, it is possible to systematically explore F-theory vacua with a fixed gauge group in the context of CICY manifolds, which is a topic that we will return to in future work \cite{usscanning}. 

\subsection{Arithmetic of the Sections}
Although the decomposition of the Picard lattice described in subsection~\ref{picsec} reveals the rank of the MW group in a systematic manner, it is a rather indirect procedure in that one still does not have explicit forms for the generating sections. In this subsection, we review what is known about the arithmetic of rational sections which, when combined with the section construction technology described in Section \ref{fibandsec}, allows us to obtain an explicit description of the MW group. 

The arithmetic of sections was derived at the level of divisor classes in Ref.~\cite{Grimm:2015wda} (see also Ref.~\cite{Morrison:2012ei} for the rank-one case), resulting in the following group law under the section addition, `$\oplus$':
\beq\label{addlaw_gen}
{\rm Div}(\sigma_1 \oplus \underbrace{\sigma_2\oplus \cdots \oplus \sigma_2}_\text{$k$ times}) = S_1 + k(S_2 - S_0) - k \pi ((S_1 - k S_0)\cdot (S_2 - S_0)) \ .
\eeq
Here, $S_m:={\rm Div}(\sigma_m)$, for $m=0,1,2$, denote the divisor classes associated to the sections $\sigma_m$ and we have chosen to identify the zero section as $\sigma_0$. The projection $\pi$ of the intersection, $D\cdot D'$, of two divisors $D$ and $D'$ in $X_n$ is defined in Ref.~\cite{Grimm:2015wda} for the $n=3, 4$ cases. In particular, for the $n=3$ case, which we will give examples of in Sections~\ref{eg2}-\ref{Sec:X3-3}, the projection is given by
\beq
\pi(D\cdot D') := (D \cdot D' \cdot D^{{\rm b}, \alpha} ) D^{\rm b}_\alpha \ , 
\eeq
where the index $\alpha=1, \cdots, h^{1,1}(B)$ is raised and lowered by the intersection matrix,
\beq
\eta_{\alpha\beta} = \hat D^{\rm b}_\alpha \cdot\hat D^{\rm b}_\beta
\eeq
of the base two-fold. For the simpler case of $n=2$, one can also show that the appropriate projection has to be defined as
\beq
\pi(D\cdot D') := (D\cdot D') D^{\rm b} \ ,
\eeq
where $D^{\rm b}$ is the pull-back of a hyperplane class in the base.  

The procedures described in Section~\ref{fibandsec} can be used to find divisor classes corresponding to true sections for a given elliptically fibered Calabi-Yau manifold. If enough sections are found this way, given the rank of the MW group, one can choose a zero section and then use the remaining ${\rm rk}\,MW(X)$ generators and the addition law~\eqref{addlaw_gen} to form a complete basis of the MW group. 

As an illustration, we return to the $K3$ example. Let us choose the zero section $\sigma_0$ to be the one we have obtained in Section~\ref{fibandsec} with the class $\cO_X(S_0)=\cO_X(-1,1,1)$. 
In Section~\ref{eg1}, it will also be shown that $\cO_X(S_1)=\cO_X(2,-1,4)$ represents another section, call it $\sigma_1$, and here we will use this fact.
The divisor classes of $k \sigma_1$ are then given by substituting $\sigma_1 \to \sigma_0$ and $\sigma_2 \to \sigma_1$ in the formula~\eqref{addlaw_gen}, 
\bea
{\rm Div}(k \sigma_1) &=& S_0 +k (S_1 - S_0 ) + k (k-1)\; \pi(S_0 \cdot ( S_1 - S_0)) \ ,  \\
&=&  (-1+3k) J_{1} + (1-2k) J_{2} + (1-11k+14k^2) J_{3} \ , 
\eea
which reproduces all the putative sections in Eq.~\eqref{gs-0}. Therefore, given that $S_0$ and $S_1$ are sections, each of those putative section classes has to also correspond to a true section, and furthermore, this proves that $\sigma_1$ fully generates the rank-one MW group. 

\vspace{0.2cm}

In the remaining sections of this paper we will demonstrate the details of the above discussions with a series of explicit examples.


\section{Example 1: A K3 example} \label{eg1}
In this section, we provide a complete analysis of the $K3$ geometry, \eqref{K3-config-0}, that we used to illustrate our general analysis in the preceding sections
\beq\label{K3-config}
\quad X_2 =\def\arraystretch{1.2}\left[\ba{c||cc} 
\IP^1_{\bold x_1} &  1 & 1 \\
\IP^2_{\bold x_2} &  1 & 2 \\
\IP^1_{\bold x_3} &  1 & 1 \\
\ea\right]  \ .
\eeq
This $K3$ surface admits an obvious genus-one fibration structure over the base $B_1=\IP^1_{\bold x_3}$, where the configuration of the fiber is given by, 
\beq\label{Fiber-config}
\quad F =\def\arraystretch{1.2}\left[\ba{c||cc} 
\IP^1_{\bold x_1} &  1 & 1 \\
\IP^2_{\bold x_2} &  1 & 2 \\
\ea\right]  \ .  
\eeq
For the purpose of giving explicit examples of results for this configuration, we choose the following generic complex structure:
\bea \nn
P_1(\bold x_1, \bold x_2, \bold x_3)&=&5 x_{1,0} x_{2,0} x_{3,0} + 9 x_{1,1} x_{2,0} x_{3,0} - 11 x_{1,0} x_{2,1} x_{3,0} + 13 x_{1,0} x_{2,2} x_{3,0} + 7 x_{1,1} x_{2,2} x_{3,0} \\ \nn
&& -\, 17 x_{1,0} x_{2,0} x_{3,1} - 17 x_{1,1} x_{2,0} x_{3,1} + 19 x_{1,0} x_{2,1} x_{3,1} - 14 x_{1,1} x_{2,1} x_{3,1} + 6 x_{1,0} x_{2,2} x_{3,1} \\  \label{K3-complex1}
&&  -\,  12 x_{1,1} x_{2,2} x_{3,1}\ , \\ \nn 
P_2(\bold x_1, \bold x_2, \bold x_3)&=&-8 x_{1,0} x_{2,0}^2 x_{3,0} - 5 x_{1,1} x_{2,0}^2 x_{3,0} - 11 x_{1,0} x_{2,0} x_{2,1} x_{3,0} + 5 x_{1,1} x_{2,0} x_{2,1} x_{3,0} + 
 7 x_{1,0} x_{2,1}^2 x_{3,0} \\ \nn
 && + \, 16 x_{1,1} x_{2,1}^2 x_{3,0} - 13 x_{1,0} x_{2,0} x_{2,2} x_{3,0}  + x_{1,1} x_{2,0} x_{2,2} x_{3,0} - 
 20 x_{1,0} x_{2,1} x_{2,2} x_{3,0} \\ \nn
 && - \, 20 x_{1,1} x_{2,1} x_{2,2} x_{3,0}  + 15 x_{1,0} x_{2,2}^2 x_{3,0} - 12 x_{1,1} x_{2,2}^2 x_{3,0} + 
 12 x_{1,0} x_{2,0}^2 x_{3,1} + 6 x_{1,1} x_{2,0}^2 x_{3,1} \\ \nn 
 && -\, 8 x_{1,0} x_{2,0} x_{2,1} x_{3,1} - 13 x_{1,1} x_{2,0} x_{2,1} x_{3,1} - 
 9 x_{1,0} x_{2,1}^2 x_{3,1} - 16 x_{1,1} x_{2,1}^2 x_{3,1} \\ \nn
 &&-\, 16 x_{1,0} x_{2,0} x_{2,2} x_{3,1} + 19 x_{1,1} x_{2,0} x_{2,2} x_{3,1} + 
 9 x_{1,0} x_{2,1} x_{2,2} x_{3,1} + 13 x_{1,1} x_{2,1} x_{2,2} x_{3,1} \\ \label{K3-complex2} 
 &&-\, 13 x_{1,0} x_{2,2}^2 x_{3,1} + 15 x_{1,1} x_{2,2}^2 x_{3,1} \ . 
\eea

\subsection{Section Analysis}
\subsubsection*{Putative Sections}

Let us start with the classification of putative sections. For a putative section $S$ labelled by $\cO_X (S)= \cO_X(b_1, b_2,b_3)$, given that the minimal base-point form integrates to unity, 
\beq
\int_{\IP^1_{\bold x_3}} J_3 = 1 \ , 
\eeq
the intersection of $S$ with the generic fiber $F_p$ is computed, as in Section \ref{putsec}, as follows.
\beq\label{k3-cond1-inter}
\int_{\IP^1_{\bold x_1} \times \IP^2_{\bold x_2} \times \IP^1_{\bold x_3}} \left[(b_1 J_1 + b_2 J_2 + b_3 J_3) \wedge J_3\right] \wedge \left[(J_1 + J_2 + J_3) \wedge (J_1 + 2J_2 + J_3) \right] = 2 b_1 + 3 b_2 \ , 
\eeq
Here $J_{1}$ and $J_{2}$ are the K\"ahler forms of $\IP^1_{\bold x_1}$ and $\IP^2_{\bold x_2}$, respectively. Therefore, the Oguiso criterion (a) demands that
\beq\label{k3-cond1}
2 b_{1} + 3 b_2 \overset{!}{=} 1 \ . 
\eeq
Via a similar intersection computation, the second criterion~\eqref{btop} leads to\footnote{For a smooth divisor $S$, the left hand side of Eq.~\eqref{k3-cond2} is $-\chi(S)$ and hence, the criterion is equivalent to $\chi(S)=\chi(B)$ unless singularities are involved.}
\beq\label{k3-cond2}
6 b_{1} b_{2} + 2b_{2}^2 + 4 b_{1} b_{3} + 6 b_{2} b_{3} \overset{!}{=} -2 \ ,
\eeq
where Eq.~\eqref{k3-cond1} has been used in simplifying the result. 
These two putative-section conditions,~\eqref{k3-cond1} and~\eqref{k3-cond2}, can be solved as
\beq\label{gs}
b_{1} = -1-3k \ , \quad b_{2} = 1+2k \ , \quad b_{3} = 1+11k+14k^2 \ . 
\eeq
with an integer parameter $k \in \IZ$. For instance, with $k= 1, 0, -1, -2$, we obtain the following putative section classes, $\cO_X(-4,3,26), ~\cO_X(-1,1,1), ~\cO_X(2,-1,4), ~\text{and~}\cO_X(5,-3,35)$, respectively.

Now the question arises as to whether the putative section classes in Eq.~\eqref{gs} indeed correspond to sections. In what follows, we will first show that two of them do by providing explicit expressions for the sections themselves (the rest will also prove to be a section later in Section~\ref{K3-section-arith}). 

\subsubsection*{Explicit Expressions for the Sections}

Let us proceed with the methodology described in Section \ref{fibandsec}. Here we will use the alternative approach discussed in the paragraph under Eq.~\eqref{equivalent} and show that it gives the same result as the method employed explicitly in that section. We will focus on the line bundle $\cO_X(S)=\cO_X ( -1, 1, 1)$, taking the solution $(b_{1}, b_{2}, b_{3}) = (-1, 1, 1)$ from the family~\eqref{gs}. The divisor class $\left[S\right] = -J_{1} + J_{2} + J_{3}$ naturally splits into two effective pieces, $S_{\rm zero}$ and $S_{\rm poles}$, such that 
\beq
\left[S_{\rm zero}\right] = J_{2} + J_{3} \ ,  \quad \left[S_{\rm pole}\right] = J_{1} \ , 
\eeq
that intersect with the generic fiber at $3$ and $2$ points, respectively (see Eq.~\eqref{k3-cond1-inter}). 
The GHS of $\cO_X(S)$ can then be constructed by appropriately choosing two GHS's, 
\beq
s_{\rm zero} \in H^0(X, \cO_X(S_{\rm zero})) \ , \quad s_{\rm pole} \in H^0(X, \cO_X({S_{\rm pole}})) \ , 
\eeq
of $\cO_X(S_{\rm zero}) = \cO_X (0,1,1)$ and $\cO_X(S_{\rm pole}) = \cO_X(1,0,0)$ so that along the generic fiber $F_p$ over $p \in B$ the two points of $F_p \cap S_{\rm pole}$ match with two of the three points of $F_p \cap S_{\rm zero}$. The unmatched point of $F_p \cap S_{\rm zero}$ should be the single intersection point of $F_p \cap S$.  

To be more concrete, let us illustrate the procedure with explicit expressions, given the complex structure in Eqs.~\eqref{K3-complex1} and~\eqref{K3-complex2}.
For a generic random choice of GHS of $\cO_X({S_{\rm pole}})=\cO_X(1,0,0)$, for example,
\beq
s_{\rm pole}(\bold x_1 , \bold x_2 , \bold x_3) = x_{1,0} + 10 x_{1,1} \ , 
\eeq
we shall look for an appropriate section of $\cO_X(S_{\rm zero})=\cO_X(0,1,1)$, 
\beq\label{szero}
s_{\rm zero} (\bold x_1 , \bold x_2 , \bold x_3) =  C_1 x_{2,0} x_{3,0}+C_2 x_{2,0} x_{3,1}+C_3 x_{2,1} x_{3,0}+C_4 x_{2,1} x_{3,1}+C_5 x_{2,2} x_{3,0}+C_6 x_{2,2} x_{3,1} \ , 
\eeq
for which $F_p \cap S_{\rm pole} \subset F_p \cap S_{\rm zero}$.
Demoting the base coordinates $\bold x_3=(x_{3,0}:x_{3,1})$ to parameters and solving the system,
\bea
P_1(\bold x_1, \bold x_2, \bold x_3) &=&0 \ ,  \\ 
P_2(\bold x_1, \bold x_2, \bold x_3) &=&0 \ , \\
s_{\rm pole} (\bold x_1, \bold x_2, \bold x_3) &=& 0 \ ,
\eea
for $\bold x_1$ and $\bold x_2$, one obtains two solutions $(\bold x_1^{(a)},\;\bold x_2^{(a)})=(\bold x_1^{(a)}(\bold x_3),\; \bold x_2^{(a)}(\bold x_3))$, for $a=1,2$.
We then substitute each of these to Eq.~\eqref{szero} and demand that $s_{\rm zero} (\bold x_1^{(a)}, \bold x_2^{(a)}) \equiv 0$ as a function of $\bold x_3$, for $a=1,2$. This turns out to fix the section $s_{\rm zero}$  uniquely (up to scaling) as
\beq\label{a1-6}
(C_1, C_2, C_3, C_4, C_5, C_6) = \lambda \; (41, -153, -110, 204, 123, 72) \ , {\text{~with~}} \lambda \in \IC \ .
\eeq 
As promised in the general discussion of Section~\ref{section-to-efib}, one can immediately confirm that this is equivalent to Eq.~\eqref{equivalent}, where the numerator is given as
\bea \nn 
-10 p_1 + p_2 &=& -10 (5 x_{2,0} x_{3,0} -11 x_{2,1}x_{3,0} + 13 x_{2,2}x_{3,0} -17 x_{2,0}x_{3,1} + 19 x_{2,1}x_{3,1} + 6 x_{2,2} x_{3,1})  \\  
&&+  (9 x_{2,0} x_{3,0} + 7 x_{2,2}x_{3,0} -17 x_{2,0}x_{3,1} -14 x_{2,1}x_{3,1} -12 x_{2,2}x_{3,1} ) \\ \nn
&=& -41 x_{2,0} x_{3,0}+153 x_{2,0} x_{3,1} + 110 x_{2,1} x_{3,0} - 204  x_{2,1} x_{3,1} -123 x_{2,2} x_{3,0} -72 x_{2,2} x_{3,1} \ , 
\eea
which agrees with Eq.~\eqref{a1-6} for $\lambda = -1$. 

Finally, having specified the two divisors, $S_{\rm zero}$ and $S_{\rm pole}$, of $X$ as the vanishing loci of $s_{\rm zero}$ and $s_{\rm pole}$, respectively,  
we can now explicitly parameterize the section, $S = S_{\rm zero} \setminus S_{\rm pole}$, of the elliptic fibration in terms of the base coordinates $\bold x_3$. We obtain the following explicit parametric expression for the section, 
\beq
x_{1,0} = A_{1,0}(\bold x_3) \ ,\quad  x_{1,1} = A_{1,1}(\bold x_3) \ ; \quad x_{2,0} = A_{2,0}(\bold x_3) \ , \quad x_{2,1}= A_{2,1}(\bold x_3) \ ,\quad x_{2,2} = A_{2,2}(\bold x_3) \ .
\eeq
Here, the polynomials $A_{1,i}(\bold x_3)$ for $i=0,1$, as well as $A_{2,i}(\bold x_3)$ for $i=0,1,2$, are, respectively, quintic and quadratic polynomials given by
{\small\bea\label{X2-1-sm}\nn 
A_{1,0} (\bold x_3) &=& 241226 x_{3,0}^5 - 2444409 x_{3,0}^4 x_{3,1} + 6970327 x_{3,0}^3 x_{3,1}^2 -  4889388 x_{3,0}^2 x_{3,1}^3 - 2858859 x_{3,0} x_{3,1}^4 + 992331 x_{3,1}^5 \ , \\\nn
A_{1,1} (\bold x_3) &=&  152844 x_{3,0}^5 - 1296506 x_{3,0}^4 x_{3,1} + 3553577 x_{3,0}^3 x_{3,1}^2 - 8289055 x_{3,0}^2 x_{3,1}^3 + 11322255 x_{3,0} x_{3,1}^4 - 5290227 x_{3,1}^5\ , \\\nn
A_{2,0} (\bold x_3) &=&77 x_{3,0}^2 - 447 x_{3,0} x_{3,1} + 144 x_{3,1}^2  \ , \\\nn
A_{2,1} (\bold x_3) &=& -82 x_{3,0}^2 - 12 x_{3,0} x_{3,1} + 306 x_{3,1}^2 \ , \\
A_{2,2} (\bold x_3) &=& -99 x_{3,0}^2 + 428 x_{3,0} x_{3,1} - 561 x_{3,1}^2 \ .
\eea}
These expressions define a rational map from the base, parameterized by ${\bf x}_3$, to the fiber, parameterized by ${\bf x}_1$ and ${\bf x}_2$. The map is well defined over every point on the base. That is, for every choice of ${\bf x}_3$ on the base manifold, we obtain a valid set of homogeneous coordinates ${\bf x}_1$ and ${\bf x}_2$ on the fiber. In particular, for no point on the base do we find that all of the homogeneous coordinates in a fiber ambient projective space factor simultaneously vanish. We thus conclude that the map is a holomorphic section, and confirm that the line bundle $\cO_X(S)=\cO_X(-1,1,1)$ is associated to this holomorphic section to the elliptic fibration. 

\vspace{0.2cm}

We can perform the same analysis for the second putative section, $\cO_X(2,-1,4)$. In order to find the associated rational map, we first need to find the rational expression for the GHS of this line bundle. With the negative degree in the $\IP^2_{\bold x_2}$, however, the naive ansatz with the linear denominator in $\bold x_2$ does not work. It turns out that the GHS can be found once we shift both the numerator and the denominator multi-degrees by $(0,2,0)$, {\it i.e.}, with the modified ansatz, 
\beq
z = \frac{N(\bold x_1, \bold x_2, \bold x_3)}{D(\bold x_2)} \ , 
\eeq
where $N \in \Gamma(X, \cO_X(2,2,4))$ and $D \in \Gamma(X, \cO_X(0,3,0))$. 
Let us take $D(\bold x_2) = x_{2,0} x_{2,1} x_{2,2}$ and tune the $90$ monomial coefficients in $N$. That is, we demand that $N$ vanishes on each of the three divisors $\{x_{2,i}=0\} \subset X$. Practically, we achieve this by substituting a sufficiently large number of points on these loci into the equation for the divisor. Furthermore, we also demand that $N$ appropriately vanishes to order $2$ on all of the points in $X$ with $\bold x_{2}=(1:0:0)$, $(0:1:0)$, or $(0:0:1)$. Then we find that only $9$ of the $90$ coefficients are undetermined. With such a tuning, $z=N/D$ is globally holomorphic. This may at first sound strange since we know that $h^0(X, \cO_X(2,-1,4))=1$. The only way to make sense of this result is that the $9$-parameter expression we have obtained for the GHS's of $\cO_X(2,-1,4)$ should only span a one-dimensional vector space of GHS. Indeed, this turns out to be the case and in the coordinate ring of $X$ they all lead to one and the same GHS up to scaling. Having specified $z=N/D$, we can now proceed to find a generic parametrization of its zero locus in terms of the base coordinates $\bold x_3$. This results in a parametric expression of the form, 
\bea
x_{1, i}= A'_{1,i}(\bold x_3, \bold x_4) &\in& \Gamma(B, \cO_B(5)) \ , \quad \text{for}\;\, i=0,1 \ , \\
x_{2, i}= A'_{2,i}(\bold x_3, \bold x_4) &\in& \Gamma(B, \cO_B(16)) \ , \quad \text{for}\;\, i=0,1,2 \ , 
\eea
where $A'_{1,i}$ coincide with the $A_{1,i}$ from Eq.~\eqref{X2-1-sm} and $A'_{2,i}$ are some fixed degree-$16$ polynomials, 
which we do not display in this paper (each of the $17$ integer coefficients has 17 to 24 digits). The map is holomorphic and we thus confirm that the line bundle $\cO_X(S)=\cO_X(2,-1,4)$ corresponds to another holomorphic section to the elliptic fibration.


%

\subsection{Locating the Singular Fibers}\label{4.2.}
\subsubsection{Weierstrass Model}\label{K3-wm}
Having proven that the unique GHS of the line bundle, 
\beq
L_z:=\cO_X(S_0) = \cO_X(-1,1,1) \ , 
\eeq
is a holomorphic section to the elliptic fibration, we choose to use $L_z$ as the bundle, which the Weierstrass coordinate $z$ is a GHS of.  
Then the other Weierstrass coordinates, $x$ and $y$, should respectively be a GHS of the following line bundles, 
\bea
L_x &:=& \cO_X (2S_0)\otimes K_B^{-2} = \cO_X(-2, 2, 6) \ , \\
L_y &:=&\cO_X(3S_0)\otimes K_B^{-3} = \cO_X(-3,3,9) \ .
\eea

In order to obtain the explicit Weierstrass model, we first need to construct an expression for the three Weierstrass coordinates, $z$, $x$, and $y$, in terms of the homogeneous coordinates, $\bold x_1$, $\bold x_2$ and $\bold x_3$.  We have already obtained an explicit expression for $z$ in the previous subsection. One can similarly choose a quadratic and a cubic polynomial in $\bold x_1$ for the denominators of global sections of $L_x$ and $L_y$, respectively and tune appropriate degree numerators to construct their GHS's. In this way, we obtain explicit rational expressions,
\bea
z&=&z(\bold x_1, \bold x_2, \bold x_3)\in \Gamma(X, L_z) \ , \\
\tilde{x}&=&\tilde{x}(\bold x_1, \bold x_2, \bold x_3)\in \Gamma(X, L_x) \ , \\
\tilde{y}&=&\tilde{y}(\bold x_1, \bold x_2, \bold x_3)\in \Gamma(X, L_y) \ , 
\eea
where $z$ is uniquely fixed up to an overall constant and $\tilde{x}$ and $\tilde{y}$ are expressed as a linear combination of $6$ and $11$ independent rational expressions, respectively. Note that the tilded variables correspond to general sections of the appropriate line bundles. The specific sections corresponding to Weierstrass coordinates will then simply be denoted as $x$ and $y$, respectively. One can independently compute the dimensions of these line bundles as
\beq
h^\bullet(X, L_z)=(1,0,0) \ , \quad h^\bullet(X, L_x)=(6,0,0) \ , \quad h^\bullet(X, L_y)=(11,0,0) \ , 
\eeq
and hence confirm that a complete basis has been obtained for each space of GHS's. 

We are now ready to construct the Weierstrass model via the procedure described in Subsection~\ref{w-procedure}. Before we start off, however, let us first convince ourselves, via cohomology numerology, that the procedure will work. In the one-dimensional space, $H^0(X, L_z)$, we find the unique $z$. Next we consider $H^0(X, L_x)$. It has $h^0(X, L_x) = 6$ independent GHS's, $5$ of which are of the form, 
\beq\label{tz}
t^{(4)}(\bold x_3) z(\bold x_1, \bold x_2, \bold x_3) \ ,
\eeq 
with $t^{(4)} \in \Gamma(B, K_B^{-2}) = \Gamma(\IP^1, \cO_{\IP^1}(4))$. The procedure tells us that there are GHS's in $\Gamma(X, L_x)$, which cannot be written in the form~\eqref{tz}, and such a GHS can be obtained as a generic linear combination of the $6$ basis elements for $\Gamma(X, L_x)$. We denote that choice by $x$. Similarly, we consider $H^0(X, L_y)$. It has $h^0(X, L_y)=11$ independent GHS's, $10$ of which are of the form, 
\beq\label{txz-tzzz}
t^{(6)}(\bold x_3)\; z(\bold x_1, \bold x_2, \bold x_3)^3  \ , \quad t^{(2)}(\bold x_3)\; z(\bold x_1, \bold x_2, \bold x_3)\; x(\bold x_1, \bold x_2, \bold x_3) \ , 
\eeq
where $t^{(6)} \in \Gamma(\IP^1, \cO_{\IP^1}(6))$ and $t^{(2)} \in \Gamma(\IP^1, \cO_{\IP^1}(2))$. Again, there exists GHS's in $\Gamma(X, L_y)$, which cannot be written in the form~\eqref{txz-tzzz}, and such a GHS can be obtained as a generic linear combination of the $11$ basis elements for $\Gamma(X, L_y)$. We denote that choice by $y$. 
Finally, let us consider $H^0(X, \cO_X(-6,6,18))$, which turns out to be $38$ dimensional. Making use of $z$, $x$, and $y$, one can construct a total of $39$ GHS's as follows:
\beq
y^2 \ , ~~ t^{(2)}yxz \ , ~~t^{(6)}yz^3  \ , ~~x^3 \ , ~~t^{(4)}x^2 z^2  \ , ~~ t^{(8)}x z^4 \ , ~~ t^{(12)}z^6  \ , 
\eeq
where $t^{(k)}\in \Gamma(\IP_{\bold x_3}^1, \cO_{\IP_{\bold x_3}^1}(k))$. Therefore, there must be a linear relation among these $39$ GHS's. This can be found by writing down a generic linear combination of the GHS's with unspecified coefficients. We then substitute a number of points on $X$ into this combination and constrain the coefficients such that the linear combination is zero on each point. If this procedure is repeated for enough points then the coefficients will be completely specified up to an overall scale. Upon an appropriate choice of that overall scale, one thus obtains the following Tate form, 
\beq
y^2 + a_1 xy z+ a_3 y z^3= x^3 + a_2 x^2 z^2+ a_4 x z^4+ a_6 z^6\ .
\eeq
In the case at hand, we obtain the explicit Tate coefficients below.
\bea\nn
a_1&=&(251-435i) x_{3,0}^2 - (479-829i) x_{3,0}x_{3,1} +(143-247i) x_{3,1}^2 \ , \\ \nn
a_2&=&(29600+ 51200i)x_{3,0}^4+ (-113000-195000i) x_{3,0}^3 x_{3,1}+ (150000+260000i) x_{3,0}^2 x_{3,1}^2 \\ \nn
&&
+ (-59100-102000i) x_{3,0} x_{3,1}^3+ (8490+14700i) x_{3,1}^4 \ , \\ \nn
a_3&=&-4250 x_{3,0}^6 -24500 x_{3,0}^5 x_{3,1}+ 3.13\times 10^6 x_{3,0}^4 x_{3,1}^2 -7.25\times 10^6 x_{3,0}^3 x_{3,1}^3+ 6.28 \times 10^6 x_{3,0}^2 x_{3,1}^4 \\ \nn 
&&
-4.33\times 10^6 x_{3,0} x_{3,1}^5+ 1.23\times 10^6 x_{3,1}^6  \ , \\ \nn
a_4&=& (352000-609000i)x_{3,0}^8+ (973000-1.69\times 10^6i)x_{3,0}^7 x_{3,1}+ (-3.49\times 10^8 + 6.05 \times 10^8 i)x_{3,0}^6 x_{3,1}^2 \\ \nn
&&
+ (1.46 \times 10^9 - 2.52 \times 10^9 i) x_{3,0}^5 x_{3,1}^3 +(-2.44\times 10^9 + 4.23 \times 10^9i) x_{3,0}^4 x_{3,1}^4 \\ \nn 
&&
+ (2.11\times 10^9 - 3.65 \times 10^9 i) x_{3,0}^3 x_{3,1}^5+ (-1.30 \times 10^9 + 2.25 \times 10^9i) x_{3,0}^2 x_{3,1}^6 \\ \nn
&&
+ (4.26\times 10^8-7.37\times 10^8i) x_{3,0} x_{3,1}^7+ (-4.63\times 10^7+8.01\times 10^7i) x_{3,1}^8 \ , \\ \nn
a_6&=& 3.61\times 10^6 x_{3,0}^{12}+ 9.68\times 10^7 x_{3,0}^{11} x_{3,1}+ 2.91\times 10^9 x_{3,0}^{10} x_{3,1}^2 -6.86\times 10^9 x_{3,0}^9 x_{3,1}^3 -1.85\times 10^{12} x_{3,0}^8 x_{3,1}^4 \\ \nn
&&
+ 8.48\times 10^{12}x_{3,0}^7 x_{3,1}^5-1.71 \times 10^{13}x_{3,0}^6 x_{3,1}^6+ 2.08\times 10^{13}x_{3,0}^5 x_{3,1}^7-1.87 \times 10^{13}x_{3,0}^4x_{3,1}^8 \\ \nn
&&
+ 1.16\times 10^{13}x_{3,0}^3x_{3,1}^9 -5.20\times 10^{12}x_{3,0}^2x_{3,1}^{10}+ 1.50\times 10^{12}x_{3,0}x_{3,1}^{11} -1.62\times 10^{11}x_{3,1}^{12} \  .
\eea
Note that here we have only reproduced the numerical coefficients in these expressions to three significant figures in order to keep the equations of a manageable size.

Via a reparameterization it is possible to obtain the Weierstrass form, 
\beq
y^2 = x^3 + f_W x z^4+ g_W z^6\ , 
\eeq 
where $f_W \in \Gamma (B, K_B^{-4}) = \Gamma(\IP^1, \cO_{\IP^1} (8))$ and $g_W \in \Gamma(B, K_B^{-6})=\Gamma(\IP^1, \cO_{\IP^1} (12))$.
The discriminant locus is then located at the vanishing of $\Delta_W := 4 f_W^3 + 27 g_W^2$, which is a homogeneous polynomial of degree $24$ in $\bold x_3 = (x_{3,0}: x_{3,1})$ and which has $24$ distinct roots. We confirm that at these $24$ points, $f_W$ and $g_W$ do not vanish and $\Delta_W$ vanishes to order $1$. Therefore, the singular fibers are of type $I_1$ from the Kodaira's classification.   

\subsubsection{Jacobian}
The configuration~\eqref{Fiber-config} describing the fiber in this example represents a complete intersection in the toric variety $\mathcal A_F = \IP_{\bold x_1}^1 \times \IP_{\bold x_2}^2$ of a bi-linear and a multi-degree $(1,2)$ polynomials. This codimension-two fiber has the PALP ID $(4,0)$, which, via the results of Ref.~\cite{Braun:2014qka}, can straightforwardly be transformed into the Jacobian as follows. The defining equations can be viewed as a function of the fiber coordinates, $\bold x_1$ and $\bold x_2$, with the base coordinates $\bold x_3$ being demoted to parameters. Then, the various monomial coefficients in the defining equations are expressed in terms of the base coordinates $\bold x_3$. The expressions for $f$ and $g$ of the Jacobian in terms of those coefficients are found from~Ref.~\cite{Braun:2014qka} with the PALP ID $(4,0)$, and thereby one obtains the Jacobian in the form, 
\beq
y^2 = x^3 + f_J x z^4+ g_J z^6\ , 
\eeq 
where $f_J \in \Gamma (B, K_B^{-4})$ and $g_J \in \Gamma(B, K_B^{-6})$. The discriminant $\Delta_J:=4 f_J^3 + 27 g_J^2$ of the Jacobian is a homogeneous polynomial of degree $24$. We confirm explicitly that this discriminant agrees with that obtained from the Weierstrass equation derived in the previous subsection:
\beq
\Delta_W \sim \Delta_J \ . 
\eeq

\subsubsection{Resolved Geometry}
Finally, we may find the discriminant locus at the level of the smooth geometry, given by Eqs.~\eqref{K3-complex1} and~\eqref{K3-complex2}. As described in Subsection~\ref{res-geom}, the fiber is singular at points in $X$ that obeys Eqs.~\eqref{fiber-sing}. In this case these equations are the two fiber-defining relations, 
\beq
P_1(\bold x_1, \bold x_2, \bold x_3)=0 \ , \quad P_2(\bold x_1, \bold x_2, \bold x_3)=0 \ , 
\eeq
together with the condition for the degeneration of the normal form, 
\beq \label{mrbadger}
{\rm d_F} P_1 \wedge {\rm d_F} P_2 = 0 \ .
\eeq
Eq.~\eqref{mrbadger} gives rise to three polynomial equations as the fiber is embedded in a threefold $\IP_{\bold x_1}^1 \times \IP_{\bold x_2}^2$, along which the exterior derivatives, ${\rm d_F}$, are taken. 
Given this system of five polynomials in $\bold x_1$, $\bold x_2$, and $\bold x_3$, we can immediately eliminate the fiber coordinates to locate the discriminant locus in the base, using Gr\"obner basis techniques. As a consequence, we obtain a single degree-$24$ polynomial, $\Delta_{\rm res}(\bold x_3)$, which can easily be seen to agree with the other two discriminant equations that we have already found:
\beq
\Delta_{\rm res} \sim \Delta_W \sim \Delta_J \ . 
\eeq 

\subsection{Arithmetic of the Sections}\label{K3-section-arith}
The sections to the elliptic fibration form an additive group. Let us take $\cO_X(-1,1,1)$ as the line bundle associated to the zero section, $\sigma_0$, and $\cO_X(2,-1,4)$ as that associated to the generator section, $\sigma_g$, of the MW group. Note that with this choice, $\sigma_g$ could potentially only generate a subgroup of the MW group. However, we will see shortly that this is not the case in this example. Denoting their divisor classes by $S_0$ and $S_g$, the divisor classes of $k \sigma_g$ are given as~\cite{Grimm:2015wda, Morrison:2012ei}
\bea \label{addlaw}
{\rm Div}(k \sigma_g) &=& S_0 +k (S_g - S_0 ) + k (k-1)\; \pi(S_0 \cdot ( S_g - S_0)) \ ,  \\
&=&  (-1+3k) J_{1} + (1-2k) J_{2} + (1-11k+14k^2) J_{3} \ .
\eea
This reproduces all of the putative sections in Eq.~\eqref{gs}. Note that the addition law in Eq.~\eqref{addlaw} is guaranteed to give true sections corresponding to multiples of $\sigma_g$ as long as the two divisors $\sigma_0$ and $\sigma_g$ are indeed both sections themselves. Since $S_0$ and $S_g$ have both been proven to be sections, we find that each of the putative sections in Eq.~\eqref{gs} is a section too, and that these sections exhaust the MW group. 

This in particular proves that the MW group is of rank $1$. The consistency of this result can be checked by analyzing the decomposition of the rank-three Picard lattice of $X$. According to the Tate-Shioda-Wazir theorem, the Picard lattice of $X$ is comprised of linearly independent basis elements of the following four types: 
\bi
\item [(1)] Base divisor classes: $B=\IP^1$ has a unique generator for the base divisor classes in this example. 
\item [(2)] Fibral divisors: As was shown by an analysis of the discriminant in the previous subsection, it turns out that there are no enhancements of fiber singularities at codimension one in this example and hence, blow-up divisors do not exist. 
\item [(3)] The zero section: $\cO_X(-1,1,1)$ has proven to lead to a (holomorphic) section, $\sigma_0$, which we may take as the zero section. 
\item [(4)] Rational sections: The arithmetic~\eqref{addlaw} of putative sections shows that the MW group has rank $1$ and is generated by $\sigma_g$, associated with the line bundle $\cO_X(2,-1,4)$. 
\ei
We find then, that the contributions from these four different types of generators indeed give rise to the Picard lattice of rank $1+0+1+1=3$, as desired in this example.

\section{Example 2: A threefold example}  \label{eg2}
We now consider an elliptic Calabi-Yau threefold example, $X_3$ with the configuration,
\beq\label{CY3-config}
\quad X_3 =\def\arraystretch{1.2}\left[\ba{c||cc} 
\IP^1_{\bold x_1} &  1 & 1 \\
\IP^2_{\bold x_2} &  1 & 2 \\
\IP^1_{\bold x_3} &  1 & 1 \\
\IP^1_{\bold x_4} & 1 & 1 \\ 
\ea\right]  \ .
\eeq
This threefold can be obtained by fibering the $K3$ surface in the previous Section over $\IP^1_{\bold x_4}$. This corresponds to the CICY \#$7675$ in the list of complete intersection Calabi-Yau threefolds~\cite{CicySymm} and has the Hodge numbers $(h^{1,1}, h^{2,1})=(4,50)$.  From the configuration matrix, an obvious genus-one fibration structure with the base $B_2 = \IP_{\bold x_3}^1 \times \IP^1_{\bold x_4}$ can be seen, where the fiber has the same configuration as Eq.~\eqref{Fiber-config},
\beq\label{conf-fiber-X3}
\quad F =\def\arraystretch{1.2}\left[\ba{c||cc} 
\IP^1_{\bold x_1} &  1 & 1 \\
\IP^2_{\bold x_2} &  1 & 2 \\
\ea\right]  \ .  
\eeq
We start by fixing a generic complex structure, for which the threefold is smooth. Instead of writing out the long expressions for the two defining equations, we will provide the list of coefficients for a given monomial ordering. Firstly, the monomial GHS's of $\cO_\cA(d_1, \cdots, d_4)$ are given the lexicographic ordering applied to the exponent list. For instance, to compare the two monomial GHS's, $m_1 = x_{1,0} x_{2,1}^2 x_{3,0} x_{4,0}$ and $m_2 = x_{1,0} x_{2,0} x_{2,2} x_{3,1} x_{4,1}$, of $\cO_\cA(1,2,1,1)$, we first read their exponents as a list of degree vectors,  
\bea
m_1 &=& \bold x_1^{(1,0)} \bold x_2^{(0,2,0)} \bold x_3^{(1,0)} \bold x_4^{(1,0)} \sim \{(1,0), (0,2,0), (1,0), (1,0)\} \ , \\
m_2 &=& \bold x_1^{(1,0)} \bold x_2^{(1,0,1)} \bold x_3^{(0,1)} \bold x_4^{(0,1)}\sim\{(1,0), (1,0,1), (0,1),(0,1) \}\ , 
\eea
then the exponent order leads to $m_2 > m_1$ and hence, $m_2$ will come earlier. With respect to this ordering for the basis monomials, our choice of the two defining equations, 
\bea
P_1&\in& \Gamma (\cA, \cO_\cA(1,1,1,1)) \ , \quad {\text{with}}\;\,h^0(\cA, \cO_\cA(1,1,1,1))=24 \ , \\
P_2&\in& \Gamma (\cA, \cO_\cA(1,2,1,1)) \ , \quad {\text{with}}\;\, h^0(\cA, \cO_\cA(1,2,1,1))=48 \ ,
\eea
are specified by the two lists of the monomial coefficients, 
\bea\label{cs-X3}
c_{P_1} &=& \{45, 4, 17, 7, 35, 69, 82, 43, 44, 46, 19, 95, 74, 100, 36, 71, 4, 43, 95, 2, 36, 9, 50, 28\} \ , \\ 
c_{P_2} &=& \{97, 60, 78, 22, 11, 27, 36, 15, 12, 53, 65, 81, 64, 12, 81, 70, 17, 9, 12, 89, 91, 43, 38, 43,\\ \nonumber & & ~6, 99, 66, 3, 58, 26, 62, 62, 43, 23, 76, 6, 40, 63, 2, 85, 42, 80, 63, 37, 75, 33, 74, 67\} \ .
\eea

\subsection{Section Analysis}
\subsubsection*{Putative Sections}

We begin by writing down the criteria that a putative section $S$, labelled by $\cO_X(S) = \cO_X(b_1, b_2, b_3, b_4)$, must satisfy. Firstly, the Oguiso criterion, that $S$ must have a single intersection point with a generic fiber, leads to 
\beq\label{X3-oguiso}
2 b_1 + 3 b_2 \overset{!}{=} 1\ .
\eeq
The base topology criteria~\eqref{btop}, one for each base divisor, $D^{\rm b}_\alpha \sim J_3, J_4$, are 
\beq\label{X3-btop}
6 b_1 b_2 + 2 b_2^2 +  4 b_1 b_\alpha + 6 b_2 b_\alpha \overset{!}{=} -2 \ , \quad {\text{for}}\;\, \alpha=3,4 \ , 
\eeq
where Eq.~\eqref{X3-oguiso} has been used in simplifying the result. 
These three equations~\eqref{X3-oguiso} and~\eqref{X3-btop} then lead to the one-parameter family of divisor classes,
\beq\label{X3-gs}
b_{1} = -1-3k \ , \quad b_{2} = 1+2k \ , \quad b_{3} = 1+11k+14k^2 \ , \quad b_4=1+11k+14k^2 \ ,
\eeq
with an integer parameter $k\in \IZ$. 
This family gives us, for the values $k=\{1,0, -1, -2\}$, the putative sections, $\cO_X(-4, 3, 26, 26)$, $\cO_X(-1,1,1,1)$, $\cO_X(2,-1,4,4)$ and $\cO_X(5, -3, 35, 35)$, respectively. 
In what follows, in order to prove that the MW group is of rank $1$, we will provide an explicit description of the associated section maps for the simplest two putative sections, 
\bea\label{ps1}
\cO_X(-1,1,1,1) \ , && {\text{with}}~~h^\bullet(X, \cO_X(-1,1,1,1))=(1,0,0,0) \ , \\ \label{ps2}
\cO_X(2,-1,4,4) \ , && {\text{with}}~~h^\bullet(X, \cO_X(2,-1,4,4))=(1,6,0,0) \ . 
\eea 

\subsubsection*{Explicit Expressions for the Sections}

Let us start with the simplest putative section, $\cO_X(-1,1,1,1)$.  We take the ansatz,
\beq\label{X3-z-gCICY}
z = \frac{N(\bold x_2, \bold x_3, \bold x_4)}{D(\bold x_1)} \ , 
\eeq
where $N \in \Gamma(X, \cO_X(0,1,1,1))$ and $D \in \Gamma(X, \cO_X(1,0,0,0))$. 
As a specific example, we may choose $D=x_{1,0}- x_{1,1}$. Then, the numerator polynomial $N$ has to be fixed up to scaling so that $z$ can be globally holomorphic. Organizing the first defining relation, $P_1$ as
\beq
P_1 = x_{1,0}\, p_0(\bold x_2, \bold x_3, \bold x_4) + x_{1,1}\, p_1(\bold x_2, \bold x_3, \bold x_4) \ , 
\eeq
where $p_0$ and $p_1$ are tri-linear polynomials, we see that on the divisor of vanishing denominator, $D(\bold x_1) = x_{1,0} - x_{1,1} = 0$, the tri-linear polynomial $p_0 + p_1$ must vanish. Therefore, upon choosing 
\bea
N &:=& p_0 +p_1 \\ \nn
&=& 119 x_{2,0} x_{3,0} x_{4,0} + 
  39 x_{2,1} x_{3,0} x_{4,0} + 
  80 x_{2,2} x_{3,0} x_{4,0} + 
  53 x_{2,0} x_{3,1} x_{4,0} + 
  177 x_{2,1} x_{3,1} x_{4,0} \\ \nn
&&   +\;   69 x_{2,2} x_{3,1} x_{4,0} + 
  104 x_{2,0} x_{3,0} x_{4,1} + 
  112 x_{2,1} x_{3,0} x_{4,1} + 
  55 x_{2,2} x_{3,0} x_{4,1} + 
  78 x_{2,0} x_{3,1} x_{4,1}  \\ \nn
&& + \;45 x_{2,1} x_{3,1} x_{4,1} + 
  123 x_{2,2} x_{3,1} x_{4,1}  \ , 
\eea
we obtain the desired GHS, the zero locus of which can easily be seen to have no singularities via the patch-wise analysis developed in Ref.~\cite{Anderson:2015iia} (note that in the above we have utilized the specific choice of complex structure that we have made in this case). 

Having specified $z=N/D$, we can now try to explicitly parameterize the zero locus in terms of the base coordinates $\bold x_3, \bold x_4$, resulting in the following generic parametrization, 
\bea\label{X3-1-sm}
x_{1, i}= A_{1,i}(\bold x_3, \bold x_4) &\in& \Gamma(B, \cO_B(5,5)) \ , \quad \text{for}\;\, i=0,1 \ , \\ \nn
x_{2, i}= A_{2,i}(\bold x_3, \bold x_4) &\in& \Gamma(B, \cO_B(2,2)) \ , \quad \text{for}\;\, i=0,1,2 \ , 
\eea
which we specify by their coefficient list for the basis monomials, as given in Appendix~\ref{erm-X3-1}.
Unlike for the $K3$ case, this section map is not holomorphic and is a rational section to the elliptic fibration. In particular, there arise $32$ points on the base $\IP^1_{\bold x_3} \times \IP^1_{\bold x_4}$ where the map is ill-defined for the $\IP^1_{\bold x_1}$ direction and $6$ points where it is ill-defined for the $\IP^2_{\bold x_2}$ direction; the latter $6$ points turn out to entirely belong to the former set. More specifically, at each of these base points, Eq.~\eqref{X3-1-sm} returns a vanishing value for all of the homogeneous coordinates of the $\IP^n$'s mentioned above. Note that the map Eq.~\eqref{X3-1-sm} gives a single point on the fiber over all of the base except for these $32$ points. The issue here is that we have naively tried to solve the system given by setting $z=0$ to find a unique value of the fiber ambient coordinates as a function of the variables in the base. At points on the base where a rational section has vertical components such a procedure can give spurious results. Let us elaborate upon this further.

Upon investigating each of the problematic base points, we observe two qualitatively different behaviors: over the $6$ base points where the rational map is ill-defined along both $\IP^1_{\bold x_1}$ and $\IP^2_{\bold x_2}$ directions, the rational section wraps a $\IP^1$ with the following configuration matrix,
\beq\label{X3-1-curvewrapping}
\def\arraystretch{1.2}\left[\ba{c||cc} 
\IP^1_{\bold x_1} &  1 & 0  \\
\IP^2_{\bold x_2} &  2 & 1 \\
\ea\right]  \ ,  
\eeq 
while over the other $26$ points it wraps a copy of $\IP^1_{\bold x_1}$. This explicitly shows that the rational section is a blowup of the base, $B=\IP^1_{\bold x_3} \times \IP^1_{\bold x_4}$ at $32$ points, which conforms with their Euler number difference,
\beq
\chi(S) - \chi(B) = 36 - 4  = 32 \ . 
\eeq

Another observation related to the above structure is as follows. Upon blowing down the smooth Calabi-Yau threefold $X$ via the contraction of the first row of its configuration matrix, one obtains a hypersurface of multi-degree $(3,2,2)$ in $\IP^2_{\bold x_2} \times \IP^1_{\bold x_3} \times \IP^1_{\bold x_4}$, which is a deformation of a smooth Calabi-Yau threefold with Euler number $-144$, while $\chi(X)$ is computed to be $-92$. The difference of these two Euler numbers is $52=2\times 26$. Thus the ambient $\IP^1$ that we are blowing down is associated with $26$ $\IP^1$'s in the Calabi-Yau geometry. This conforms with the fact that a degeneration associated with this $\IP^1$ in the section map was associated with exactly $26$ $\IP^1$'s above specific points in the base.

\vspace{0.2cm}

Let us now move on to analyzing the second putative section, $\cO_X(2,-1,4,4)$. In order to find the explicit map, we first need to find the rational expression for the GHS of this line bundle. With the negative degree in the $\IP^2_{\bold x_2}$, however, the naive ansatz with the linear denominator in $\bold x_2$ does not work. It turns out that the GHS can be found once we shift both the numerator and the denominator multi-degrees by $(0,2,0,0)$, {\it i.e.}, with the modified ansatz, 
\beq
z = \frac{N(\bold x_1, \bold x_2, \bold x_3, \bold x_4)}{D(\bold x_2)} \ , 
\eeq
where $N \in \Gamma(X, \cO_X(2,2,4,4))$ and $D \in \Gamma(X, \cO_X(0,3,0,0))$. 
As a choice of denominator, we may take $D(\bold x_2) = x_{2,0} x_{2,1} x_{2,2}$ and tune the $450$ monomial coefficients in $N$. That is, we demand that $N$ vanishes on each of the three divisors $\{x_{2,i}=0\} \subset X$, in practice employing the same techniques as we have in previous sections. Then, only $129$ coefficients of the $450$ remain free. This does not guarantee that the rational form tuned as such is regular yet, since the denominator may vanish to order $2$ on the loci in $X$ with $\bold x_{2}=(1:0:0)$, $(0:1:0)$, or $(0:0:1)$. Interestingly, we find that $N$ also vanishes to order $2$ on such loci. We therefore conclude that the tuned form is indeed regular and corresponds to GHS's. This may at first sound strange since we know that $h^0(X, \cO_X(2,-1,4,4))=1$. The only way to make sense of this result is that the $129$-parameter expression we have obtained for the GHS's of $\cO_X(2,-1,4,4)$ should only span a one-dimensional vector space, and it indeed turns out that in the coordinate ring of $X$ they all lead to one and the same GHS up to scaling. 

Having specified $z=N/D$ for this section, we can now proceed to find a generic parametrization of its zero locus in terms of the base coordinates $\bold x_3, \bold x_4$, resulting in the parametric expression of the form, 
\bea \label{thelad}
x_{1, i}= A'_{1,i}(\bold x_3, \bold x_4) &\in& \Gamma(B, \cO_B(5,5)) \ , \quad \text{for}\;\, i=0,1 \ , \\
x_{2, i}= A'_{2,i}(\bold x_3, \bold x_4) &\in& \Gamma(B, \cO_B(16,16)) \ , \quad \text{for}\;\, i=0,1,2 \ .
\eea
Here $A'_{1,i}$ coincide with the $A_{1,i}$ from Eq.~\eqref{X3-1-sm} and $A'_{2,i}$ are some fixed polynomials, which we do not display in this paper (each of the $289$ integer coefficients has 27 to 37 digits). Equipped with such an explicit map, one can show that this defines a rational section to the elliptic fibration. The map Eq.~\eqref{thelad} is ill-defined in the same manner as for the previous GHS in this section for the $\IP^1_{\bold x_1}$ direction at $32$ points on the base $\IP^1_{\bold x_3} \times \IP^1_{\bold x_4}$ and for the $\IP^2_{\bold x_2}$ direction at $92$ points on the base, the former $32$ points entirely belonging to the latter set. As before, this corresponds to the GHS giving rise to a rational section wrapping $\IP^1$'s in the fiber over some points on the base.

\subsection{Locating the Singular Fibers}
All the steps in this subsection are a straightforward analogue of those in Subsection~\ref{4.2.}. We will thus be brief here and will mainly state the results of the analysis. 

\subsubsection{Weierstrass Model} 
Having proven that the unique GHS of the line bundle, 
\beq
L_z:=\cO_X(S_0) = \cO_X(-1,1,1,1) \ , 
\eeq
is a rational section to the elliptic fibration, we choose to use $L_z$ as the bundle, which the Weierstrass coordinate $z$ is a GHS of, and take 
\bea
L_x &:=& \cO_X (2S_0)\otimes K_B^{-2} = \cO_X(-2, 2, 6, 6) \ , \\
L_y &:=&\cO_X(3S_0)\otimes K_B^{-3} = \cO_X(-3,3,9,9) \ ,
\eea
to which the other Weierstrass coordinates, $x$ and $y$, are associated.

We first need the explicit expressions for the three Weierstrass coordinates $z$, $x$ and $y$ in terms of the homogeneous coordinates $\bold x_{r=1, \cdots, 4}$. We have already performed this computation for $z$ in the previous subsection. Using similar techniques to those outlined already we can choose a quadratic and a cubic polynomial in $\bold x_1$ for the denominator and tune the numerator, to construct the other GHS's, $x$ and $y$, respectively. In this manner, we obtain explicit rational expressions,
\bea
z&=&z(\bold x_1, \bold x_2, \bold x_3, \bold x_4)\in \Gamma(X, L_z) \ , \\
x&=&x(\bold x_1, \bold x_2, \bold x_3, \bold x_4)\in \Gamma(X, L_x) \ , \\
y&=&y(\bold x_1, \bold x_2, \bold x_3, \bold x_4)\in \Gamma(X, L_y) \ , 
\eea
where $z$ is uniquely fixed up to an overall constant, while $x$ and $y$ are expressed as a linear combination of $26$ and $59$ independent rational expressions, respectively. One can independently compute the cohomology dimensions of these line bundles as
\beq
h^\bullet(X, L_z)=(1,0,0,0) \ , \quad h^\bullet(X, L_x)=(26,32,0,0) \ , \quad h^\bullet(X, L_y)=(59, 128, 0,0) \ ,
\eeq
and hence confirm that a complete basis has been obtained for each space of GHS's. 

We are then ready to construct the Weierstrass model via the procedure of Subsection~\ref{w-procedure}. The cohomology numerology works out as in previous examples, with a GHS of $L_x$ not being proportional to $z^2$ and a GHS of $L_y$ not being constructed from terms proportional to $z^3$ or $x z$. The Weierstrass relation itself is a section of $L_w :=L_y^{\otimes 2}$ whose zeroth cohomology is one less in dimension than the space that can be spanned by monomials in $x,y,z$ and the base coordinates of the correct degree, as expected. 

As in the $K3$ case, one can find a unique linear relation among the $335$ GHS's of $L_w$, using the same techniques that have been employed in that case. Upon an appropriate rescaling, one thus obtains the Tate form, 
\beq
y^2 + a_1 xy z+ a_3 y z^3= x^3 + a_2 x^2 z^2+ a_4 x z^4+ a_6 z^6\ , 
\eeq
with the explicit Tate coefficients $a_i$, and via the reparameterization, gets to the Weierstrass form, 
\beq
y^2 = x^3 + f_W x z^4+ g_W z^6\ , 
\eeq 
where $f_W \in \Gamma (B, K_B^{-4}) = \Gamma(\IP^1\times \IP^1, \cO_{\IP^1} (8,8))$ and $g_W \in \Gamma(B, K_B^{-6})=\Gamma(\IP^1\times \IP^1, \cO_{\IP^1} (12,12))$.
The discriminant locus is then located at the vanishing of $\Delta_W := 4 f_W^3 + 27 g_W^2$, which is a homogeneous polynomial of multi-degree $(24,24)$ in $(\bold x_3, \bold x_4)$. The expression for the Tate form and the discriminant are rather large and so we do not reproduce them here. The only explicit information we will require going forward with this example is that, given the discriminant polynomial $\Delta_W$, we confirm that all the codimension-one singularities are of type $I_1$. 

\subsubsection{Jacobian}
The configuration~\eqref{conf-fiber-X3} of the elliptic fiber is the same as the $K3$ case and is a complete intersection in the toric variety $\mathcal A_F = \IP_{\bold x_1}^1 \times \IP_{\bold x_2}^2$, with the PALP ID $(4,0)$. Via the results of Ref.~\cite{Braun:2014qka}, it can straightforwardly be transformed into the Jacobian as follows. The defining equations can be viewed as a function of the fiber coordinates, $\bold x_1$ and $\bold x_2$, with the base coordinates $\bold x_3$ and $\bold x_4$ demoted to parameters. Then, the various monomial coefficients in the defining equations are expressed in terms of the base coordinates $\bold x_3$ and $\bold x_4$. Using the expressions for $f$ and $g$ of the Jacobian in terms of those coefficients, one obtains the Jacobian in the form, 
\beq
y^2 = x^3 + f_J x z^4+ g_J z^6\ , 
\eeq 
where $f_J \in \Gamma (B, K_B^{-4})$ and $g_J \in \Gamma(B, K_B^{-6})$. The discriminant $\Delta_J:=4 f_J^3 + 27 g_J^2$ of the Jacobian is a homogeneous polynomial of bi-degree $(24, 24)$ and we confirm that it agrees with the Weierstrass form:
\beq
\Delta_W \sim \Delta_J \ . 
\eeq

\subsubsection{Resolved Geometry}
Finally, we can explore the discriminant locus at the level of the smooth geometry associated with the choice of complex structure in Eq.~\eqref{cs-X3}. Elliptic fibers admit a singularity at points in $X$ obeying Eqs.~\eqref{fiber-sing}, which, much like the $K3$ case, are the defining relations for the fiber,  
\beq\label{PQ-X3}
P_1=0 \ , \quad P_2=0 \ , 
\eeq
together with the degeneration condition, 
\beq\label{dPdQ-X3}
{\rm d_F} P_1 \wedge {\rm d_F} P_2 = 0 \ ,
\eeq
where the latter gives rise to three polynomial equations as the fiber is embedded in a threefold $\IP_{\bold x_1}^1 \times \IP_{\bold x_2}^2$.
Given this system of five polynomials in $\bold x_r$, we could in principle eliminate the fiber coordinates as in the $K3$ case to obtain a single polynomial $\Delta_{\rm res} \in \Gamma(B, K_B^{-12})$ for the discriminant locus embedded in $B$. However, the elimination process never finishes in a reasonable amount of time. In order to compare this discriminant locus with the previous ones, we take a numerical analysis based on a set of generic points. That is, first we choose a large number of random points that solve the system~\eqref{PQ-X3} and~\eqref{dPdQ-X3} and substitute them into the previously obtained polynomial, $\Delta_W \sim \Delta_J$ to confirm that it vanishes on all of those points. This shows that the discriminant obtained from the smooth geometry is a subset of that obtained from the Weierstrass and Jacobian forms. Similarly by choosing random points on the base that solve $\Delta_W =0 (=\Delta_J)$ and substituting them into Eqs.~\eqref{PQ-X3} and~\eqref{dPdQ-X3} we can show that the discriminant obtained from the Weierstrass form is a subset of that obtained from the analysis of the resolved geometry. In this manner, we can conclude that
\beq
\Delta_{\rm res} \sim \Delta_W \sim \Delta_J \ . 
\eeq 

\subsection{Arithmetic of the Sections}

We have so far obtained two legitimate sections, associated with $\cO_X(-1,1,1,1)$ and $\cO_X(2,-1,4,4)$, respectively. Let us take $\cO_X(-1,1,1,1)$ as the line bundle for a zero section, $\sigma_0$, and $\cO_X(2,-1,4,4)$ for the generator section, $\sigma_g$, of (potentially a subgroup of) the MW group. Denoting their divisor classes by $S_0$ and $S_g$, the divisor classes of $k \sigma_g$ are given by applying Eq.~\eqref{addlaw_gen} as
\bea\label{addlaw_X3-1}
{\rm Div}(k \sigma_g) &=& S_0 +k (S_g - S_0 ) + k (k-1)\; \pi(S_0 \cdot ( S_g - S_0)) \ ,  \\
&=&  (-1+3k) J_{1} + (1-2k) J_{2} + (1-11k+14k^2) J_{3} +(1-11k+14k^2) J_{4} \ ,
\eea
which reproduces all the putative sections in Eq.~\eqref{X3-gs}. Note that the addition law guarantees to give the divisor classes of multiples of $\sigma_g$ in the MW group. Given that $S_0$ and $S_g$ have both proven to be a section, we learn that each of the putative sections in Eq.~\eqref{X3-gs} is a section, too, and that there are no more sections.

As in the previous example, we can check what we have learned about this geometry for consistency by analyzing the splitting of the rank-four Picard lattice of $X$. The four different types of the divisor classes of $X$ are as follows:
\bi
\item [(1)] Base divisor classes: $B=\IP^1 \times \IP^1$ has two independent generators.
\item [(2)] Fibral divisors: We have seen that there are no enhancement of fiber singularities at codimension one and hence, appropriate vertical divisors do not exist.
\item [(3)] Zero section: $\cO_X(1,-1,1,1)$ has proven to be a rational section, which we may take as the zero section.
\item [(4)] Rational sections: It has also been shown in the previous paragraph that the MW group has rank $1$.
\ei
Then, the contributions from these four different types of generators indeed give rise to the correct rank, $2+0+1+1=4$, as desired.


\section{Example 3: A threefold example with non-abelian symmetry} \label{Sec:X3-2}
In the two simple examples we have so far looked at, the discriminant locus is irreducible and the singular fibers are only of type $I_1$. For a more non-trivial example with enhanced fiber singularities, in this Section, we analyze another threefold case with the configuration,
\beq\label{X3-2-conf}
\quad X_3 =\def\arraystretch{1.2}\left[\ba{c||cccccc} 
\IP^1_{\bold x_1} &  1 & 1 & 0 & 0 & 0 & 0 \\
\IP^1_{\bold x_2} &  1 & 0 & 1 & 0 & 0 & 0 \\
\IP^2_{\bold x_3} &  0 & 2 & 0 & 0 & 0 & 1 \\
\IP^3_{\bold x_4} & 0 & 0 & 1 & 1 & 1 & 1 \\
\IP^1_{\bold x_5} & 1 & 0 & 0 & 1 & 0 & 0 \\
\IP^1_{\bold x_6} & 0 & 1 & 0 & 0 & 1 & 0 \\ 
\ea\right]  \ .
\eeq
This case corresponds to the CICY $\#5075$ of the CICY threefold list \cite{CicySymm} and has the Hodge numbers $(h^{1,1}, h^{2,1}) = (7, 30)$. Note in particular that the complete intersection is not ``favorable'' in that $h^{1,1}(X) = h^{1,1}(\mathcal A) +1$. However, much of our analysis of the sections and fiber types are insensitive to favorability and we can still apply our techniques to this configuration. There also exists, as with almost all CICY threefolds, a nested Calabi-Yau fibration structure of the form,
\beq
X_3 \quad\longrightarrow\quad B_2 = \IP^{1}_{\bold x_5} \times \IP^1_{\bold x_6} \quad\longrightarrow\quad B_1 = \IP^1_{\bold x_6} \ , 
\eeq
and hence can be explored for the Heterotic/F-theory duality (see Ref.~\cite{phys}). Note that the elliptic fibration has the fiber with the following configuration, 
\beq\label{X3-2-fiber}
\quad F =\def\arraystretch{1.2}\left[\ba{c||cccccc} 
\IP^1_{\bold x_1} &  1 & 1 & 0 & 0 & 0 & 0 \\
\IP^1_{\bold x_2} &  1 & 0 & 1 & 0 & 0 & 0 \\
\IP^2_{\bold x_3} &  0 & 2 & 0 & 0 & 0 & 1 \\
\IP^3_{\bold x_4} & 0 & 0 & 1 & 1 & 1 & 1 \\
\ea\right]  \ . 
\eeq 
We start by fixing a generic complex structure, for which the threefold is smooth. We specify our choice of the six polynomials, $P_1$-$P_6$, each by a list of monomial coefficients, again, in the lexicographic order for the monomial exponents that was introduced in Section \ref{eg2}: 
\bea
c_{P_1} &=& \{20, 6, 8, 19, 6, 17, 18, 14 \} \ , \\
c_{P_2} &=& \{8, 2, 6, 18, 6, 3, 5, 8, 8, 19, 13, 15, 15, 11, 18, 20, 6, 3, 14, 2, 16, 17, 9, 11 \} \ , \\
c_{P_3} &=& \{2, 7, 10, 18, 7, 19, 14, 9 \} \ , \\
c_{P_4} &=& \{ 7, 5, 3, 19, 3, 1, 17, 2\} \ , \\
c_{P_5} &=& \{ 10, 1, 11, 19, 13, 4, 15, 19\} \ , \\
c_{P_6} &=& \{ 4, 3, 11, 20, 19, 4, 8, 17, 7, 7, 12, 16\} \ . 
\eea

\subsection{Section Analysis}

\subsubsection*{Putative Sections}

A putative section $S$, which we label as $\cO_X(S) = \cO_X(b_1, \cdots, b_6)$, must have a single topological intersection with a generic fiber, leading to
\beq\label{X3-2-oguiso}
2 b_1+ 2 b_2 + 3 b_3 + 2 b_4 \overset{!}{=} 1 \ .
\eeq
In addition, we have the base topology criteria~\eqref{btop}, one for each base divisor, $D_\alpha^{\rm b} \sim J_5, J_6$, which give the following conditions, 
\bea\label{X3-2-btop}
&&6 b_1 b_3 + 
 6 b_2 b_3 + 2 
b_3^2+ 
 4 b_1 b_4 + 
 4 b_2 b_4 + 
 8 b_3 b_4  +  2 
b_4^2+ 
 4 b_1 b_6 + 
 4 b_2 b_6 \\ \nn &&+ 
 6 b_3 b_6 + 4 b_4 b_6 \overset{!}{=} -2 \ ,  \\ \nn &&
  4 b_1 b_2 + 
 8 b_1 b_3 + 
 6 b_2 b_3 + 2 
b_3^2+ 
 8 b_1 b_4 +
 4 b_2 b_4 + 
 8 b_3 b_4 + 2 
b_4^2+ 
 4 b_1 b_5 + 
 4 b_2 b_5  \\ \nn && + 
 6 b_3 b_5 + 4 b_4 b_5 \overset{!}{=} -2 \ ,
\eea
where Eq.~\eqref{X3-2-oguiso} has been used for a simplification. These three equations~\eqref{X3-2-oguiso} and~\eqref{X3-2-btop} then lead to the following three-parameter family of divisor classes, 
\bea\label{X3-2-ps-param}
b_1 &=& -1-3k_1-k_2-k_3 \ , \\ \nn
b_2 &=& k_2\ , \\  \nn
b_3 &=& 1+2k_1\ , \\ \nn
b_4 &=& k_3\ ,  \\ \nn
b_5 &=& 2+16k_1+20k_1^2 + 3 k_2 + 8 k_1 k_2 + 2 k_2^2 + 4 k_3 + 12 k_1 k_3 + 4 k_2 k_3 + 3 k_3^2 \ , \\ \nn
b_6 &=& 1+ 11 k_1 + 14 k_1^2 + k_3 + 4 k_1 k_3 + k_3^2 \ , 
\eea
with $k_1, k_2, k_3 \in \IZ$, which, for some small parameter values, $(k_1, k_2, k_3) = (0,0,-1), (0,-1,0), (0,0,0)$, for example, give the putative sections, $\cO_X(0,0,1,-1,1,1)$, $\cO_X(0,-1,1,0,1,1)$, $\cO_X(-1,0,1,0,2,1)$, respectively, all of which turn out to have the line-bundle cohomologies,
\beq\label{1000}
h^{\bullet}(X, \cO_X(S)) = (1,0,0,0) \ .
\eeq 
We must now confirm that these putative sections do indeed correspond to genuine sections of the fibration.

\subsubsection*{Explicit Expressions for the Sections} We will analyze the simplest putative section associated with $L_z:=\cO_X(0,0,1,-1,1,1)$, whose GHS of the form,
\beq\label{X3-2-z}
z= \frac{N(\bold x_3, \bold x_5, \bold x_6)}{D(\bold x_4)}  \in \Gamma(X, L_z) \ , 
\eeq
is sought for, where $N \in \Gamma(X, \cO_X(0,0,1,0,1,1))$ and $D \in \Gamma(X, \cO_X(0,0,0,1,0,0))$. 
We choose $D=x_{4,0}$ and  proceed to determine the numerator polynomial by combining the last three defining equations, $P_4$-$P_6$. We can write these without loss of generality as,
\beq
P_4=\sum\limits_{i=0}^3 x_{4,i} \, u_{i}(\bold x_5) \ , \quad P_5=\sum\limits_{i=0}^3 x_{4,i}\, v_{i}(\bold x_6) \ , \quad P_6=\sum\limits_{i=0}^3 x_{4,i} \, w_{i}(\bold x_3) \ , 
\eeq
where $u_i$, $v_i$ and $w_i$ are linear in $\bold x_5$, $\bold x_6$ and $\bold x_3$, respectively. Then, on the divisor $\{x_{4,0}=0\} \subset X$, we have 
\beq
\sum\limits_{i=1}^3 x_{4,i} \, u_{i}(\bold x_5) = 0 \ , \quad \sum\limits_{i=1}^3 x_{4,i}\, v_{i}(\bold x_6) = 0 \ , \quad \sum\limits_{i=1}^3 x_{4,i} \, w_{i}(\bold x_3) =0  \ .
\eeq
Therefore, with the denominator $D=x_{4,0}$, if the numerator polynomial is chosen as
\bea
N&:=& -u_3 v_2 w_1 + u_2 v_3 w_1 + u_3 v_1 w_2 - u_1 v_3 w_2 - u_2 v_1 w_3 + u_1 v_2 w_3 \\  \label{X3-2-N}
&=& 
1154 x_{3,0} x_{5,0} x_{6,0}+534 x_{3,1} x_{5,0} x_{6,0}+568 x_{3,2} x_{5,0} x_{6,0}+1794 x_{3,0} x_{5,1} x_{6,0} \\ \nn
&&+\,
1864 x_{3,1} x_{5,1} x_{6,0}+543 x_{3,2} x_{5,1} x_{6,0}+1993 x_{3,0} x_{5,0} x_{6,1}+1319 x_{3,1} x_{5,0} x_{6,1} \\  \nn
&&+\,
2395 x_{3,2} x_{5,0} x_{6,1}-2380 x_{3,0} x_{5,1} x_{6,1}-1571 x_{3,1} x_{5,1} x_{6,1}-2887 x_{3,2} x_{5,1} x_{6,1}  \ , 
\eea
the rational expression~\eqref{X3-2-z} for $z$ is the desired GHS, the zero locus of which can easily be seen to have no singularities via a direct calculation. With this expression, we obtain an explicit parametrization of its zero locus of the form,
\bea\label{X3-2-sm}
x_{1,i}=A_{1,i}(\bold x_5, \bold x_6) &\in& \Gamma(B, \cO_B(2,3)) \ , \quad {\text{for}}~i=0,1 \ , \\ \nn
x_{2,i}=A_{2,i}(\bold x_5, \bold x_6) &\in& \Gamma(B, \cO_B(3,3)) \ , \quad {\text{for}}~i=0,1 \ , \\ \nn
x_{3,i}=A_{3,i}(\bold x_5, \bold x_6) &\in& \Gamma(B, \cO_B(1,1)) \ , \quad {\text{for}}~i=0,1, 2 \ , \\ \nn
x_{4,i}=A_{4,i}(\bold x_5, \bold x_6) &\in& \Gamma(B, \cO_B(4,4)) \ , \quad {\text{for}}~i=0,1, 2, 3 \ , 
\eea 
which we present in Appendix~\ref{erm-X3-2} for concreteness. 
These parameterizations provide a rational section to the elliptic fibration. In particular, the map is ill-defined, in a similar manner to what has been seen in previous sections, for the $\IP^3_{\bold x_4}$ direction at $22$ points on the base $\IP^1_{\bold x_5} \times \IP^1_{\bold x_6}$. Amongst these $22$ points, for $15$ the map is ill-defined also for the $\IP^2_{\bold x_3}$ direction, for $9$ the map is ill-defined also for the $\IP^1_{\bold x_1}$ and the $\IP^2_{\bold x_3}$ directions, and for $1$ point the map is ill defined for all four ambient projective-space directions. Note that the Euler number difference,
\beq
\chi(S) - \chi(B) = 26- 4= 22 \ ,
\eeq
indicates that the rational section wraps a $\IP^1$ in the fiber over these $22$ base points. 

One can perform similar computations for the other two small putative sections, $\cO_X(0,-1,1,0,1,1)$, $\cO_X(-1,0,1,0,2,1)$. Interestingly, both of them lead to exactly the same section map as in Eqs.~\eqref{X3-2-sm}. However, they should be understood as vertical divisors attached to the genuine section and provide the first explicit examples of putative sections failing to be a section, even when the strong cohomology condition in Eq.~\eqref{1000} holds. For more detailed analysis of this geometry, we refer the readers to Appendix~\ref{rps}. 

\subsection{Locating the Singular Fibers}
It is possible to apply all three of the different techniques for analyzing singular fibers that we have discussed in this paper to this case. However, in practice, the procedure to compute the Weierstrass model in larger cases is rather expensive in terms of computational resources. The time cost is also high in analyzing the resolved geometry; eliminating the fiber coordinates from the criterion for the fiber singularity to obtain the hypersurface equation for the discriminant locus in the base $B_2$ is a Gr\"obner basis calculation and therefore scales badly with the size of the problem. On the other hand, the Jacobian in this case can be analyzed quickly, upon blowing the geometry down until the codimension of the elliptic fiber becomes less than or equal to $2$. Therefore, for this example, we only present our exploration of Jacobian to analyze the singular fibers.

Since the fiber configuration~\eqref{X3-2-fiber} has codimension $6$, Jacobian of $X$ cannot be obtained by the prescription in Refs.~\cite{art, Braun:2011ux, Braun:2014qka}. We therefore first blow the geometry down by a chain of contractions along fiber directions as described in Section \ref{mdelajac}. Let us consider the following chain of three blow downs along the fiber directions, $\bold \IP^1_{\bold x_{1}}$, $\IP^1_{\bold x_{2}}$ and $\IP^3_{\bold x_4}$, in turn: 
\beq\label{chain1}
\def\arraystretch{1.2}
\left[\ba{c||cccccc} 
\IP^1_{\bold x_1} &  1 & 1 &0&0&0&0\\
\IP^1_{\bold x_2} &  1 & 0 &1&0&0&0\\
\IP^2_{\bold x_3} &  0 & 2 &0&0&0&1\\
\IP^3_{\bold x_4} &  0 & 0 &1&1&1&1\\
\IP^1_{\bold x_5} &  1 & 0 &0&1&0&0\\
\IP^1_{\bold x_6} &  0 & 1 &0&0&1&0\\
\ea\right]  \rightarrow
\def\arraystretch{1.2}\left[\ba{c||ccccc} 
\IP^1_{\bold x_2} &  1  &1&0&0&0\\
\IP^2_{\bold x_3} &  2 &0&0&0&1\\
\IP^3_{\bold x_4} &  0 &1&1&1&1\\
\IP^1_{\bold x_5} &  1 &0&1&0&0\\
\IP^1_{\bold x_6} &  1 &0&0&1&0\\
\ea\right] \rightarrow
\def\arraystretch{1.2}\left[\ba{c||cccc} 
\IP^2_{\bold x_3} &  2&0&0&1\\
\IP^3_{\bold x_4} &  1&1&1&1\\
\IP^1_{\bold x_5} &  1 &1&0&0\\
\IP^1_{\bold x_6} &  1 &0&1&0\\
\ea\right] \rightarrow
\def\arraystretch{1.2}\left[\ba{c||c} 
\IP^2_{\bold x_3} &  3\\
\IP^1_{\bold x_5} &  2\\
\IP^1_{\bold x_6} &  2\\
\ea\right] \ , 
\eeq
to get the fiber blown down to a cubic hypersurface in $\IP^2_{\bold x_{3}}$, with which we can immediately form the Jacobian of the form,
\beq
y^2 = x^3 + f_J x z^4+ g_Jz^6 \ , 
\eeq 
where $f_J \in \Gamma (B, K_B^{-4})$ and $g_J \in \Gamma(B, K_B^{-6})$. The discriminant $\Delta_J:=4 f_J^3 + 27 g_J^2$ of the Jacobian is a homogeneous polynomial of bi-degree $(24, 24)$. Upon analyzing the singular fibers of the resolved geometry, without eliminating the fiber coordinates to obtain the discriminant locus in the base, we have numerically compared the two discriminant loci and have shown that they are the same:
\beq
\Delta_J \sim \Delta_{\rm res} \ . 
\eeq
Note that there are five more chains of blow downs that lead to a cubic fibration by performing contractions in different orders. We find that all of these lead to the same Jacobian.  

\subsection{Discriminant Locus Analysis}

An explicit analysis of the discriminant locus $\Delta_J$ in this example reveals that it has the following factorization structure, 
\beq\label{factorize}
\Delta_{24, 24} (\bold x_5, \bold x_6) = (F^{(1)}_{1,0}(\bold x_5, \bold x_6))^2\; (F^{(2)}_{1,0}(\bold x_5, \bold x_6))^2 \; (F^{(3)}_{1,1}(\bold x_5, \bold x_6))^2 \; F^{(4)}_{18,22} (\bold x_5, \bold x_6)\ .
\eeq
Here, the subscript pairs denote the bi-degrees. The vanishing locus $L = \{\Delta=0\} \subset B$ thus decomposes accordingly as
\beq
L = \bigcup\limits_{I=1}^4 L^{(I)} \ , 
\eeq
where $L^{(I)} = \{F^{(I)}=0\} \subset B$ are the vanishing loci of the individual factors in Eq.~\eqref{factorize}. 
\vskip 1cm
\begin{figure}[h]
\centering
\includegraphics[scale=0.16]{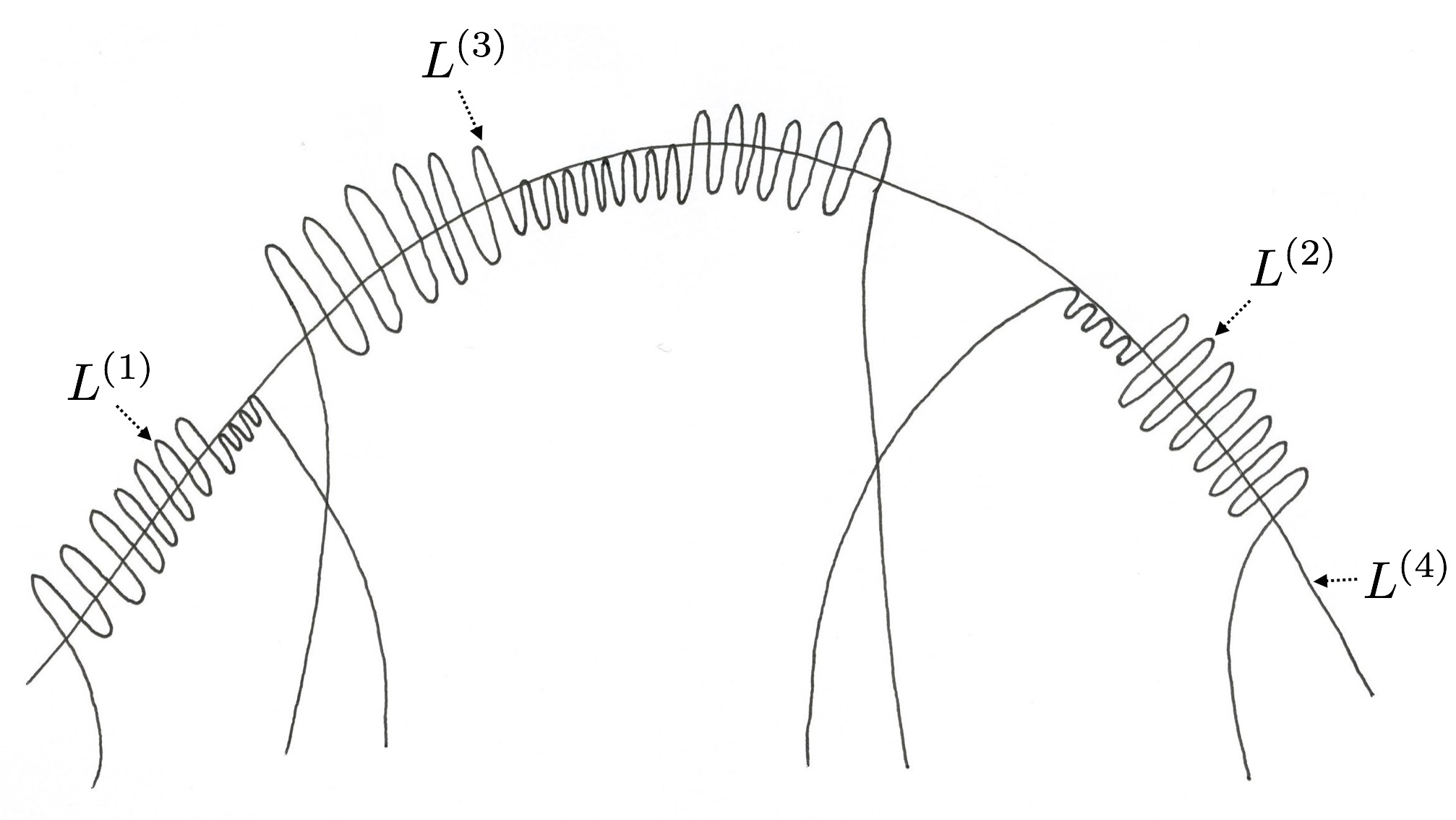}
\caption{\footnotesize Schematic picture of the discriminant locus, $L \subset B = \IP^1_{\bold x_5} \times \IP^1_{\bold x_6}$, where $\Delta(\bold x_5, \bold x_6)$ vanishes. The locus decomposes to four pieces $L^{(I)}$, $I=1, 2, 3, 4$. }\label{f:disc-locus}
\end{figure}
\vskip 1cm
By further analyzing the vanishing orders of $f$, $g$ and $\Delta$, we see the following singularity structure and the enhancement pattern (characterized by Kodaira type \cite{kodaira}):
\begin{itemize}
\item $L^{(I)}$, for $I=1,2,3$, are of type $I_2$ and $L^{(4)}$ is of type $I_1$;
\item $|L^{(I)} \cap L^{(3)}|=1$, for $I=1,2$, and at each of the two points the singularity enhances to $I_4$; 
\item $|L^{(I)} \cap L^{(4)}|=18$, for $I=1,2$, and at 14 points the singularity enhances to $I_3$ and at the remaining 4 points (each with multiplicity 2) to $III$;
\item $|L^{(3)} \cap L^{(4)}|=32$ and at 24 points the singularity enhances to $I_3$ and at the remaining 8 points (each with multiplicity 2) to $III$.
\end{itemize}
The decomposition of the discriminant locus is schematically depicted in Fig.~\ref{f:disc-locus}. 

\subsection{Arithmetic of the Sections} 
Since we have presented a complete section analysis in several other cases, in this example we will simply content ourselves with an analysis of the decomposition of the rank-seven Picard lattice of $X$. The four different types of the divisor classes of $X$ are as follows: 
\bi
\item [(1)] Base divisor classes: $B=\IP^1 \times \IP^1$ have two independent generators, 
\bea\label{X3-2-(1)}
&&\cO_X(0,0,0,0,1,0) \ , \\ \nn 
&&\cO_X(0,0,0,0,0,1) \ . 
\eea 
\item [(2)] Fibral divisors: as we have seen, there are three codimension-one loci of singularity enhancement to $I_2$ and hence, three vertical/blow-up divisors. For completeness, we provide a choice of three independent vertical-divisor classes,
\bea\label{X3-3-(2)}
&&\cO_X(-1,1,0,0,1,0) \ , \\ \nn 
&&\cO_X(1,-1,0,0,1,0) \ , \\ \nn
&&\cO_X(0,-1,0,1,0,0) \ , 
\eea 
each of which is effective with a unique GHS. Note that they do not meet with a generic fiber (when substituted into the left hand side of Eq.~\eqref{X3-2-oguiso} we obtain zero). 
\item [(3)] Zero section: $\cO_X(0,0,1,-1,1,1)$ is a rational section, which we may take as the zero section. 
\item [(4)] Rational sections: In order for the above six independent divisor classes of three different types, together with the rational sections, to span the rank-seven Picard lattice, there must exist an additional rational section generating a $MW$ group of rank one. Given that the other six divisor classes, all obtained from the favorable sector, generate the rank-six sub-lattice, the Mordell-Weil generator must lie (in part) in the non-favorable sector of the Picard lattice, which we will not discuss in this paper. Nevertheless, the decomposition of the Picard lattice alone is enough to demonstrate that such a group of sections exists. 
\ei

\section{Example 4: A threefold example with higher rank Mordell-Weil group} \label{Sec:X3-3}
In this Section, we will present an example of an elliptic CICY threefold with ${\rm rk}\, MW(X) = 4$. The configuration of this geometry (CICY \#7907 of the CICY threefold list \cite{CicySymm}) is given by
\beq\label{X3-3-conf}
\quad X_3 =\def\arraystretch{1.2}\left[\ba{c||cccccccc} 
\IP^2_{\bold x_1} &  1 & 0 & 1 & 0 & 0 & 0 & 1 & 0 \\
\IP^2_{\bold x_2} &  0 & 1 & 0 & 1 & 0 & 0 & 0 & 1 \\
\IP^3_{\bold x_3} & 0 & 0 & 0 & 0 & 1 & 1 & 1 & 1 \\
\IP^1_{\bold x_4} & 0 & 0 & 1 & 1 & 0 & 0 & 0 & 0 \\ 
\IP^1_{\bold x_5} & 1 & 1 & 0 & 0 & 0 & 0 & 0 & 0 \\
\IP^1_{\bold x_6} &  0 & 0 & 0 & 0 & 1 & 1 & 0 & 0 \\
\IP^1_{\bold x_7} &  0 & 0 & 0 & 0 & 2 & 0 & 0 & 0 \\
\ea\right]
\quad \longrightarrow \quad B_2 = \IP^1_{\bold x_6}\times \IP^1_{\bold x_6} \ , 
\eeq
where the elliptic fibers have the following configuration, 
\beq\label{X3-3-fiber}
\quad F =\def\arraystretch{1.2}\left[\ba{c||cccccccc} 
\IP^2_{\bold x_1} &  1 & 0 & 1 & 0 & 0 & 0 & 1 & 0 \\
\IP^2_{\bold x_2} &  0 & 1 & 0 & 1 & 0 & 0 & 0 & 1 \\
\IP^3_{\bold x_3} & 0 & 0 & 0 & 0 & 1 & 1 & 1 & 1 \\
\IP^1_{\bold x_4} & 0 & 0 & 1 & 1 & 0 & 0 & 0 & 0 \\ 
\IP^1_{\bold x_5} & 1 & 1 & 0 & 0 & 0 & 0 & 0 & 0 \\
\ea\right] \ . 
\eeq
It has the Hodge numbers $(h^{1,1}, h^{2,1})=(7,30)$. We start our analysis by first choosing a generic complex structure, described by the following monomial coefficients listed in the lexicographic order for the monomial exponents described in Section \ref{eg2}, 
\bea
c_{P_1} &=& \{18,14,14,13,14,13\} \ , \\ \nn
c_{P_2} &=& \{18,16,19,13,9,15\} \ , \\ \nn
c_{P_3} &=& \{2,20,7,16,2,15 \} \ , \\ \nn
c_{P_4} &=& \{4,15,3,2,5,19\} \ , \\ \nn
c_{P_5} &=& \{6,7,12,20,4,15,15,2,16,20,2,1,4,15,19,14,10,10,6,9,16,19,7,16\} \ , \\ \nn
c_{P_6} &=& \{11,4,10,7,11,8,17,9\} \ ,  \\ \nn
c_{P_7} &=& \{10,7,5,7,1,13,20,7,3,13,18,3\} \ , \\ \nn
c_{P_8} &=& \{3,4,6,6,20,4,17,3,4,20,20,20\} \ . 
\eea

\subsection{Section Analysis}
\subsubsection*{Putative Sections}
A putative section $S$, which we label as $\cO_X(S) = \cO_X(b_1, \cdots, b_6)$, must have a single topological intersection with a generic fiber, leading to
\beq\label{X3-3-oguiso}
3b_1 + 3b_2 + 2b_3 + 2b_4 + 2 b_5 \overset{!}{=} 1 \ , 
\eeq
and the base topology criteria~\eqref{btop}, one for each base divisor, $D_\alpha^{\rm b} \sim J_6, J_7$, give the following conditions, 
{
\bea\label{X3-3-btop}\nn
&&
2 
b_1^2 + 
 8 b_1 b_2 + 2 
b_2^2+ 
 12 b_1 b_3 + 
 12 b_2 b_3 + 4 
b_3^2 + 
 4 b_1 b_4 + 
 4 b_2 b_4 + 
 8 b_3 b_4 +   
 4 b_1 b_5  \\ 
 &&+~ 
 4 b_2 b_5 + 
 8 b_3 b_5 + 
 4 b_4 b_5 + 
 6 b_1 b_\alpha + 
 6 b_2 b_\alpha + 
 4 b_3 b_\alpha + 
 4 b_4 b_\alpha + 4 b_5 b_\alpha \overset{!}{=} -2 \ , 
\eea}
for $\alpha=6,7$, where Eq.~\eqref{X3-3-oguiso} has been used for a simplification. These two equations,~\eqref{X3-3-oguiso} and~\eqref{X3-3-btop}, then lead to the following four-parameter ($k_{m} \in \IZ$, for $m=1,2,3,4$) family of divisor classes.
\bea\label{X3-3-ps-param}
b_1 &=& -1-2k_1 - k_2\ , \\ \nn
b_2 &=& k_2\ , \\  \nn
b_3 &=& k_3\ , \\ \nn
b_4 &=& k_4\ ,  \\ \nn
b_5&=&2 + 3 k_1 - k_3 - k_4 \ , \\ \nn
b_6 &=& 2 + 10 k_1 + 8 
k_1^2 + 2 k_2 + 
 4 k_1 k_2 + 2 
k_2^2 - 4 k_3 - 
 4 k_1 k_3 + 2 
k_3^2 - 4 k_4 - 
 6 k_1 k_4 + 
 2 k_3 k_4 + 2 
k_4^2 \ , \\ \nn
b_7 &=& 2 + 10 k_1 + 8 
k_1^2 + 2 k_2 + 
 4 k_1 k_2 + 2 
k_2^2 - 4 k_3 - 
 4 k_1 k_3 + 2 
k_3^2 - 4 k_4 - 
 6 k_1 k_4 + 
 2 k_3 k_4 + 2 
k_4^2 \ .
\eea
With some small parameter values, this family of putative sections gives rise to the following possibilities:
\beq\label{smallparams}
\def\arraystretch{1.2}\ba{|c|c|c|} 
\hline 
S~&(k_1, k_2, k_3, k_4)~&~\cO_X(S) \\ \hline 
S_0&(0, -1, 0, 1) & \cO_X(0, -1, 0, 1, 1, 0, 0)  \\ 
S_1&(0, -1, 1, 0 ) & \cO_X( 0, -1, 1, 0, 1, 0, 0) \\ 
S_2&(0, -1, 1, 1 ) & \cO_X(0, -1, 1, 1, 0, 0, 0 ) \\ 
S_3&(-1, 0, 0, -1 ) & \cO_X(1, 0, 0, -1, 0, 0, 0 ) \\ 
S_4 
&(0, 0, 0, 1 ) & \cO_X(-1, 0, 0, 1, 1, 0, 0 ) \\ 
\hline
\ea \ , 
\eeq
all of which turn out to have the line-bundle cohomologies,
\beq
h^{\bullet}(X, \cO_X(S)) = (1,0,0,0) \ .
\eeq 
We must next ensure that these putative sections do correspond to true sections of the fibration. 

\subsubsection*{Explicit Expressions for the Sections} 

We will fully analyze the putative section associated with the first line bundle in Eq.~\eqref{smallparams}, $L_z:=\cO_X(S_0)=\cO_X(0,-1,0,1,1,0,0)$, whose GHS is of the form,
\beq\label{X3-3-z}
z= \frac{N(\bold x_4, \bold x_5)}{D(\bold x_2)}  \in \Gamma(X, L_z) \ .
\eeq
Here $N \in \Gamma(X, \cO_X(0,0,0,1,1,0,0))$ and $D \in \Gamma(X, \cO_X(0,1,0,0,0,0,0))$. 
We choose $D=x_{2,0}$ and consider the second and the fourth defining equations, 
\bea
P_2 &=&x_{2,0} q_0 (\bold x_5) + x_{2,1} q_1 (\bold x_5) + x_{2,2} q_2(\bold x_5)  \ , \\ 
P_4 &=&x_{2,0} u_0 (\bold x_4) + x_{2,1} u_1 (\bold x_4) + x_{2,2} u_2(\bold x_4)   \ ,  
\eea
where $q_{i}$ and $u_{i}$, for $i=0,1,2$, are linear in $\bold x_5$ and $\bold x_4$, respectively. Then, on the divisor $\{x_{2,0}=0\} \subset X$, we have 
\bea
x_{2,1} q_1 (\bold x_5) + x_{2,2} q_2(\bold x_5) &=& 0 \ , \\ 
x_{2,1} u_1 (\bold x_4) + x_{2,2} u_2(\bold x_4) &=&0 \ , 
\eea
and hence, with the choice, 
\beq
N:=q_1 u_2 - q_2 u_1 =  68 x_{4, 0} x_{5, 0} + 
 343 x_{4, 1} x_{5, 0} + 
 20 x_{4, 0} x_{5, 1} + 
 217 x_{4, 1} x_{5,1}\ , 
\eeq
the rational expression~\eqref{X3-3-z} for $z$ is the desired GHS. The zero locus of this GHS can easily be seen to have no singularities via a patch-wise analysis of the system. 
With this expression, we obtain the explicit parameterization of its zero locus of the divisor of the form,
\bea\label{X3-3-sm}
x_{1,i}=A_{1,i}(\bold x_6,\bold x_7) &\in& \Gamma(B, \cO_B) \ , \quad {\text{for}}~i=0,1,2\ , \\ \nn
x_{2,i}=A_{2,i}(\bold x_6,\bold x_7) &\in& \Gamma(B, \cO_B(2,2)) \ , \quad {\text{for}}~i=0,1,2 \ , \\ \nn
x_{3,i}=A_{3,i}(\bold x_6,\bold x_7) &\in& \Gamma(B, \cO_B(2,2)) \ , \quad {\text{for}}~i=0,1,2,3 \ , \\ \nn
x_{4,i}=A_{4,i}(\bold x_6,\bold x_7) &\in& \Gamma(B, \cO_B) \ , \quad {\text{for}}~i=0,1 \ , \\ \nn
x_{5,i}=A_{5,i}(\bold x_6,\bold x_7) &\in& \Gamma(B, \cO_B) \ , \quad {\text{for}}~i=0,1 \ ,  
\eea 
which we present in Appendix~\ref{erm-X3-3} for concreteness. 
These parameterizations provide a rational section to the elliptic fibration. In particular, the section map is ill-defined, in the same manner as has been seen in preceding sections. This phenomenon occurs for the $\IP^2_{\bold x_2}$ direction at $8$ points on the base $\IP^1_{\bold x_6} \times \IP^1_{\bold x_7}$, amongst which are $4$ points where the map is ill-defined also for the $\IP^3_{\bold x_3}$ direction. Note that the Euler number difference,
\beq
\chi(S) - \chi(B) = 12- 4= 8 \ ,
\eeq
indicates that the rational section indeed wraps a $\IP^1$ in the fiber over these $8$ base points.

Similarly, one can analyze the other smooth, rational sections, $S_{m=1,2,3,4}$, using the same procedures. We thereby obtain the explicit rational maps for each of those MW generators and also confirm the way in which fibral $\IP^1$'s are wrapped by each rational section. 

\subsection{Locating the Singular Fibers}
As in Section~\ref{Sec:X3-2}, exploration of the Jacobian is the simplest way in which to obtain the information we need about the singular fibers. With the fiber~\eqref{X3-3-fiber} being of codimension $8$, to easily read off the Jacobian of $X$ we first need to blow the geometry down. Let us consider the following chain of three contractions along the $\bold \IP^2_{\bold x_{1}}$, $\IP^3_{\bold x_{3}}$, and $\bold \IP^1_{\bold x_{4}}$ directions, in turn: 
\bea\label{chain1-X3-3}
\def\arraystretch{1.2}\left[\ba{c||cccccccc} 
\IP^2_{\bold x_1} &  1 & 0 & 1 & 0 & 0 & 0 & 1 & 0 \\
\IP^2_{\bold x_2} &  0 & 1 & 0 & 1 & 0 & 0 & 0 & 1 \\
\IP^3_{\bold x_3} & 0 & 0 & 0 & 0 & 1 & 1 & 1 & 1 \\
\IP^1_{\bold x_4} & 0 & 0 & 1 & 1 & 0 & 0 & 0 & 0 \\ 
\IP^1_{\bold x_5} & 1 & 1 & 0 & 0 & 0 & 0 & 0 & 0 \\
\IP^1_{\bold x_6} &  0 & 0 & 0 & 0 & 1 & 1 & 0 & 0 \\
\IP^1_{\bold x_7} &  0 & 0 & 0 & 0 & 2 & 0 & 0 & 0 \\
\ea\right]&\rightarrow&
\def\arraystretch{1.2}\left[\ba{c||cccccc} 
\IP^2_{\bold x_2} &  0&1 & 1 & 0 & 0 & 1  \\
\IP^3_{\bold x_3} & 1&0 & 0 & 1 & 1 & 1  \\
\IP^1_{\bold x_4} & 1&0 & 1 & 0 & 0 & 0  \\ 
\IP^1_{\bold x_5} & 1&1 & 0 & 0 & 0 & 0  \\
\IP^1_{\bold x_6} &  0&0 & 0 & 1 & 1 & 0  \\
\IP^1_{\bold x_7} &  0&0 & 0 & 2 & 0 & 0  \\
\ea\right] \\ \nonumber &\rightarrow&
\def\arraystretch{1.2}\left[\ba{c||cccccc} 
\IP^2_{\bold x_2} &  1&1& 1  \\
\IP^1_{\bold x_4} & 1&0&1  \\ 
\IP^1_{\bold x_5} & 1&1 & 0  \\
\IP^1_{\bold x_6} &  2&0& 0  \\
\IP^1_{\bold x_7} & 2 & 0 & 0  \\
\ea\right] \rightarrow
\def\arraystretch{1.2}\left[\ba{c||cccccc} 
\IP^2_{\bold x_2} &  2& 1  \\
\IP^1_{\bold x_5} & 1&1   \\
\IP^1_{\bold x_6} &  2& 0  \\
\IP^1_{\bold x_7} & 2  & 0  \\
\ea\right]  \;.
\eea
After this procedure the fiber has been blown down to a complete intersection of two hypersurfaces in $\IP^2_{\bold x_{2}} \times \IP^1_{\bold x_{5}}$ of degrees $(2,1)$ and $(1,1)$, respectively. This blown-down fiber has the PALP ID $(4,0)$ and, based on the results of Ref.~\cite{Braun:2014qka}, we can read the Jacobian of the form,
\beq
y^2 = x^3 + f_J x z^4+ g_Jz^6 \ , 
\eeq 
where $f_J \in \Gamma (B, K_B^{-4})$ and $g_J \in \Gamma(B, K_B^{-6})$. The discriminant $\Delta_J:=4 f_J^3 + 27 g_J^2$ of the Jacobian is a homogeneous polynomial of degree $(24,24)$.
Upon going through the analysis of the resolved geometry, without eliminating the fiber coordinates, we are able to numerically compare the the two discriminant loci and confirm that
\beq
\Delta_J \sim \Delta_{\rm res} \ . 
\eeq
Having obtained the explicit discriminant polynomial $\Delta_J$, one can proceed to analyze its factorization properties and we find that all the codimension-one fibers are of type $I_1$. 

\subsection{Arithmetic of the Sections} 
Let us denote sections corresponding to $S_m$, $m=0, 1, \cdots, 4$ in \eqref{smallparams}, by $\sigma_m \in MW (X)$. Here we will show that the MW group is of rank $4$, and that the lattice is generated by these $\sigma_m$. Specifically, based on the group law~\eqref{addlaw_gen} at the level of divisor classes, we will show, upon choosing $\sigma_0$ to be the zero section, the following: 
\bi
\item $\sigma_{1}, \cdots, \sigma_{4}$ span a lattice of rank $4$;
\item $\sigma_1, \cdots, \sigma_4$ generate all the putative section classes in the family~\eqref{X3-3-ps-param}. 
\ei
These two combined will prove not only that $\sigma_{m=1,2,3,4}$ is a basis of the MW group, but also that all of the putative section classes in Eq.~\eqref{X3-3-ps-param} correspond to a genuine section. 
Both of the bullet points above can be shown by forming the divisor class of a most general linear combination of $\sigma_m$. Upon sequentially applying the addition law~\eqref{addlaw_gen}, one sees that
\beq\label{lc-form}
{\rm Div}(\bigoplus\limits_{m=1}^4 l_m \sigma_m) = \sum_{r=1}^{7}{\beta_r} J_r \ , 
\eeq
where the coefficients $\beta_r$ are given in terms of the $l_m$ as
\bea\label{lc-detail}
\beta_1&=&l_3 - l_4 \ , \\ \nn
\beta_2&=& -1 + l_3 + l_4\ , \\ \nn
\beta_3&=&l_1 + l_2 \ , \\ \nn
\beta_4&=&1 - l_1 - 2 l_3 \ , \\ \nn
\beta_5&=&1 - l_2 - l_3 \ , \\ \nn
\beta_6&=& -2 l_1 + 2 
l_1^2 - 2 l_2 + 
 2 l_1 l_2 + 2 
l_2^2 - 2 l_3 + 
 2 l_1 l_3 + 2 
l_3^2 - 2 l_4 + 2 
l_4^2\ , \\ \nn
\beta_7&=& -2 l_1 + 2 
l_1^2 - 2 l_2 + 
 2 l_1 l_2 + 2 
l_2^2 - 2 l_3 + 
2l_1 l_3 + 2
l_3^2- 2 l_4 + 2 
l_4^2\ . \\ \nn
\eea
With Eqs.~\eqref{lc-form} and~\eqref{lc-detail}, one can immediately see that 
\beq
\bigoplus\limits_{m=1}^4 l_m \sigma_m = \sigma_0 \quad \Rightarrow \quad {\rm Div}(\bigoplus\limits_{m=1}^4 l_m \sigma_m) = S_0 \quad \Rightarrow \quad l_m = 0,~~\text{for}~m=1, 2, 3, 4\ , 
\eeq
which proves the first bullet point. In addition, the $b_r$ from Eq.~\eqref{X3-3-ps-param} equal $\beta_r$ for $r=1, \cdots, 7$, with the following choice of $l_m$,
\beq
l_1 = 1 + 2 k_1 - k_4 \,; ~\quad l_2 = -1 - 2 k_1 + k_3 + k_4 \,; ~\quad l_3 = - k_1 \,; ~\quad l_4 =  1 + k_1 + k_2 \ , 
\eeq
which proves the second bullet point.
Therefore, we have in particular shown ${\rm rk}\, MW(X) = 4$, which is consistent with the following decomposition of the independent divisor classes:
\bi
\item [(1)] Base divisor classes: $B=\IP^1 \times \IP^1$ has two independent generators, 
\bea\label{X3-3-(1)}
&&\cO_X(0,0,0,0,0,1,0) \ , \\ \nn 
&&\cO_X(0,0,0,0,0,0,1) \ . 
\eea 
\item [(2)] Fibral divisors: As we have seen, there do not arise any enhancement of fiber singularities at codimension one and hence, no blow-up divisors either. 
\item [(3)] Zero section: $\sigma_0$ has been proven to be a rational section, which we may take as the zero section with the divisor class, 
\bea\label{X3-3-(3)}
&&\cO_X(0,-1,0,1,1,0,0) \ . 
\eea
\item [(4)] Rational sections: $MW(X)$ has been proven to be generated by $\sigma_m$, $m=1,2,3,4$, with the respective divisor classes, 
\bea\label{X3-3-(4)}
&&\cO_X(0,-1,1,0,1,0,0) \ , \\ \nn
&&\cO_X(0,-1,1,1,0,0,0) \ , \\ \nn
&&\cO_X(1,0,0,-1,0,0,0) \ ,  \\ \nn
&&\cO_X(-1,0,0,1,1,0,0) \ . 
\eea
\ei
Then, the contributions from these four different types of generators give rise to the correct rank for the Picard lattice, $2+0+1+4=7$. 
Furthermore, the seven divisor classes in Eqs.~\eqref{X3-3-(1)},~\eqref{X3-3-(3)} and~\eqref{X3-3-(4)} indeed span the entire Picard lattice and hence form a basis.


\section*{Acknowledgements}
The authors would like to thank A.~Grassi, A.~Kapfer, T.~Pantev, W.~Taylor for helpful conversations. The work of LA (and XG in part) is supported by NSF grant PHY-1417337 and that of JG (and SJL in part) is supported by NSF grant PHY-1417316. This project is part of the working group activities of the 4-VA initiative ``A Synthesis of Two Approaches to String Phenomenology".


\begin{appendix}
\section{Reducible Putative Sections and Non-Flat Fibers}\label{rps}
In this section, we provide examples of reducible putative sections that decompose into a genuine section and a vertical divisor. Let us consider the CICY threefold discussed in Section~\ref{Sec:X3-2}, with the elliptic fibration given by the CICY and the fiber configuration matrices~\eqref{X3-2-conf} and~\eqref{X3-2-fiber}, respectively,
\beq
\quad X_3 =\def\arraystretch{1.2}\left[\ba{c||cccccc} 
\IP^1_{\bold x_1} &  1 & 1 & 0 & 0 & 0 & 0 \\
\IP^1_{\bold x_2} &  1 & 0 & 1 & 0 & 0 & 0 \\
\IP^2_{\bold x_3} &  0 & 2 & 0 & 0 & 0 & 1 \\
\IP^3_{\bold x_4} & 0 & 0 & 1 & 1 & 1 & 1 \\
\IP^1_{\bold x_5} & 1 & 0 & 0 & 1 & 0 & 0 \\
\IP^1_{\bold x_6} & 0 & 1 & 0 & 0 & 1 & 0 \\ 
\ea\right] ; \quad   
F =\def\arraystretch{1.2}\left[\ba{c||cccccc} 
\IP^1_{\bold x_1} &  1 & 1 & 0 & 0 & 0 & 0 \\
\IP^1_{\bold x_2} &  1 & 0 & 1 & 0 & 0 & 0 \\
\IP^2_{\bold x_3} &  0 & 2 & 0 & 0 & 0 & 1 \\
\IP^3_{\bold x_4} & 0 & 0 & 1 & 1 & 1 & 1 \\
\ea\right]  \ ,  
\eeq
with the base $B=\IP^1_{\bold x_5} \times \IP^1_{\bold x_6}$. Recall that in Section~\ref{Sec:X3-2} we have identified a family of putative sections (see Eq.~\eqref{X3-2-ps-param}), from which three line bundles with small degrees, 
\beq
L_1=\cO_X(0,0,1,-1,1,1)\ , \quad L_2=\cO_X(0,-1,1,0,1,1) \ , \quad L_3=\cO_X(-1,0,1,0,2,1) \ ,
\eeq
were chosen for discussion. Note that all these three line bundles have the cohomology, 
\beq
h^\bullet (X, L_i) = (1,0,0,0)\ , \quad {\text{for}}~ i=1,2,3.  
\eeq
The unique global holomorphic section of $L_1$ is of the form, 
\beq
s_1=\frac{N(\bold x_3, \bold x_5, \bold x_6)}{x_{4,0}} \ , 
\eeq
with the numerator polynomial $N$ given in Eq.~\eqref{X3-2-N}, and its vanishing locus has proven smooth, leading to a genuine section to the elliptic fibration (see Eq.~\eqref{X3-2-sm} and Appendix~\ref{erm-X3-2}). On the other hand, the other two putative sections associated with the line bundles, $L_2$ and $L_3$, are neither smooth nor irreducible, while leading to the same section map as the $L_1$ case. As claimed in the main text, they should be thought of as the genuine section attached with an additional vertical divisor. Since the two cases are similar in nature, we will only present a full analysis for one of them, $L_2$. We start with the following injective mapping for cohomology groups, 
\beq
H^0(X, L_1) \times H^0(X, L) \longrightarrow H^0(X, L_2) \ , 
\eeq 
where $L:=\cO_X(0,-1,0,1,0,0)$ satisfies $L_1 \otimes L = L_2$ and $h^0(X, L)=1$. 
Given such an injection, the unique GHS of $L_2$ should be of the form, 
\beq
s_2 = s_1 \, s \ , 
\eeq
where $s$ is the GHS of $L$ that can be written, for instance, as
\beq
s=\frac{r_1(\bold x_4)}{x_{2,0}}\ , 
\eeq
with $r_1(\bold x_4)$ read from the expansion of the third defining equation for $X$, 
\beq
P_3 = x_{2,0} r_0(\bold x_4) + x_{2,1} r_1(\bold x_4) \ . 
\eeq

The divisor $\{s=0\} \subset X$ does not intersect with a generic fiber and is in fact vertical. One way to see this is to compute the left hand side of the Oguiso criterion \eqref{intersection-c} and notice that one obtains zero. 
Via numerical algebraic geometry techniques, one can further check that the singular locus of the putative section associated with $L_2$ is of dimension $1$, which is consistent with the fact that the genuine section and the vertical divisor meets at a curve. 

Similarly, the relevant injection for the $L_3$ case is,
\beq
H^0(X, L_1) \times H^0(X, L') \longrightarrow H^0(X, L_3) \ , 
\eeq
where $L':=\cO_X(-1,0,0,1,1,0)$ satisfies $L_1 \otimes L' = L_3$ and $h^0(X, L')=1$. The argument for this case follows along exactly analogous lines.  

\vspace{0.2cm}

The decomposition structure described above is rather clean. For some configurations, however, one may face a more exotic situation where the elliptic fiber is non-flat and a point on the base pulls back to a vertical divisor. 
As an illustration, let us consider the CICY threefold (with CICY \#5075 \cite{CicySymm}), in which an elliptic fibration can be found with the CICY and the fiber configurations,
\beq\label{rps-Efib-2}
\quad X_3 =\def\arraystretch{1.2}\left[\ba{c||cccccc} 
\IP^1_{\bold x_1} &  1 & 0 & 1 & 0 &  0& 0  \\
\IP^1_{\bold x_2} &  0 & 1 & 0 & 0 & 1 & 0 \\
\IP^2_{\bold x_3} &  0 & 2 & 0 & 0 & 0 & 1 \\
\IP^3_{\bold x_4} &  0 & 0 & 1 & 1 & 1 & 1 \\
\IP^1_{\bold x_5} &  1 & 1 & 0 & 0 & 0 & 0 \\
\IP^1_{\bold x_6} &  1 & 0 & 0 & 1 & 0 & 0 \\
\ea\right] ; \quad   
F =\def\arraystretch{1.2}\left[\ba{c||cccccc} 
\IP^1_{\bold x_1} &  1 & 0 & 1 & 0 &  0& 0  \\
\IP^1_{\bold x_2} &  0 & 1 & 0 & 0 & 1 & 0 \\
\IP^2_{\bold x_3} &  0 & 2 & 0 & 0 & 0 & 1 \\
\IP^3_{\bold x_4} &  0 & 0 & 1 & 1 & 1 & 1 \\
\ea\right]  \ ,  
\eeq
over the base $B=\IP^1_{\bold x_5}\times \IP^1_{\bold x_6}$. Note that this is the same CICY geometry as the one analyzed in Section~\ref{Sec:X3-2}, with the rows appropriately interchanged ({\it i.e.}, the total space is the same but we are examining a different fibration structure here). One can easily confirm, via the simple topological checks of Section \ref{section-to-efib}, that the two line bundles, 
\beq
L_1=\cO_X(1,0,1,-1,0,1) \ , \quad L_2=\cO_X(0,0,1,-1,1,2) \ , 
\eeq
correspond to putative sections. Furthermore, they both have the cohomology, 
\beq
h^\bullet(X, L_i)=(1,0,0,0) \ , \quad {\text{for}}~i=1,2\ . 
\eeq
Let us expand the first, the third, the fourth, and the sixth defining equations for $X$ as
\bea \label{p1lad}
P_1&=& \sum\limits_{i=0}^1 x_{1,i} \, p_{i}(\bold x_5, \bold x_6) \ , \\
P_3&=& \sum\limits_{i=0}^3 x_{4,i} \, r_{i}(\bold x_1) \ , \\ 
P_4&=& \sum\limits_{i=0}^3 x_{4,i} \, u_{i}(\bold x_6) \ , \\ 
P_6&=& \sum\limits_{i=0}^3 x_{4,i} \, w_{i}(\bold x_3) \ , 
 \eea
where $r_{i}$, $u_{i}$, and $w_i$, for $i=0,1,2,3$, are all linear in their respective variables and $p_{i}$, for $i=0,1$, are bilinear in $\bold x_5$ and $\bold x_6$. Then, the GHS of $L_1$ can be constructed as
\beq
s_1 = \frac{-r_3 u_2 w_1 + r_2 u_3 w_1 + r_3 u_1 w_2 - r_1 u_3 w_2 - r_2 u_1 w_3 + r_1 u_2 w_3}{x_{4,0}} \ , 
\eeq
and its zero locus can easily be proven smooth. 
Then, exactly the same steps as those used for all the examples in the main text can be applied to show that this putative section is a genuine section to the elliptic fibration~\eqref{rps-Efib-2}. 

For the GHS of $L_2$, we consider the injection, 
\beq
H^0(X, L_1) \times H^0(X, L) \longrightarrow H^0(X, L_2) \ , 
\eeq
where $L:=\cO_X(-1,0,0,0,1,1)$ satisfies $L_1 \otimes L = L_2$ and $h^0(X, L) = 1$. The GHS $s_2$ of $L_2$ then factors as 
\beq
s_2 = s_1 \, s \ , 
\eeq
where $s$ is the section of $L$ that can be written, for instance, as
\beq
s=\frac{p_1(\bold x_5, \bold x_6)}{x_{1,0}} \sim - \frac{p_0(\bold x_5, \bold x_6)}{x_{1,1}}\ . 
\eeq
Thus, the putative section $\{s_2=0\} \subset X$ decomposes into the genuine section $\{s_1=0\}$ and the vertical divisor $\{s=0\}$. However, the vertical divisor in this case is the non-flat fiber over the two points in the base with $p_0(\bold x_5, \bold x_6) = 0 = p_1(\bold x_5, \bold x_6)$\footnote{In order to see that this fiber is non-flat at these points note that, when these two terms vanish, the first defining equation in the fiber configuration in Eq.~\eqref{rps-Efib-2} becomes trivial (see Eq.~\eqref{p1lad})}.

It is interesting to look in more detail at the structure associated to these two points. In order to do so, we proceed to analyze the Jacobian of the blown-down geometry, obtained via the following chain of contractions:
\beq\nn
\def\arraystretch{1.2}\left[\ba{c||cccccccc} 
\IP^1_{\bold x_1} &  1 & 0 & 1 & 0 &  0& 0  \\
\IP^1_{\bold x_2} &  0 & 1 & 0 & 0 & 1 & 0 \\
\IP^2_{\bold x_3} &  0 & 2 & 0 & 0 & 0 & 1 \\
\IP^3_{\bold x_4} &  0 & 0 & 1 & 1 & 1 & 1 \\
\IP^1_{\bold x_5} &  1 & 1 & 0 & 0 & 0 & 0 \\
\IP^1_{\bold x_6} &  1 & 0 & 0 & 1 & 0 & 0 \\
\ea\right]\rightarrow
\def\arraystretch{1.2}\left[\ba{c||cccccc} 
\IP^1_{\bold x_2} &  0 & 1 & 0 &  1 & 0 \\
\IP^2_{\bold x_3} &  0 & 2 & 0 &  0 & 1 \\
\IP^3_{\bold x_4} &  1 & 0 & 1 & 1 & 1 \\
\IP^1_{\bold x_5} &  1 & 1 & 0 & 0 & 0 \\
\IP^1_{\bold x_6} &  1 & 0 & 1 & 0 & 0 \\
\ea\right] \rightarrow
\def\arraystretch{1.2}\left[\ba{c||cccccc} 
\IP^2_{\bold x_3} &  2 & 0 &  0 & 1 \\
\IP^3_{\bold x_4} &  1 & 1 & 1 &  1 \\
\IP^1_{\bold x_5} &  1 & 1 & 0 &  0 \\
\IP^1_{\bold x_6} &  0 & 1 & 1 &  0 \\
\ea\right] \rightarrow
\def\arraystretch{1.2}\left[\ba{c||cccccc} 
\IP^2_{\bold x_3} &  3 \\
\IP^1_{\bold x_5} &  2 \\
\IP^1_{\bold x_6} &  2 \\ 
\ea\right] \ . 
\eeq
The Jacobian of the resulting cubic fibration \cite{art} has a Weierstrass form whose $f_J$, $g_J$, and $\Delta_J$, at these two base points, vanish to order $4$, $6$, and $12$, respectively. Conversely, one may analyze non-generic codimension-two points as follows. The $\Delta_J$ turns out to factorize as 
\beq
\Delta_{24,24}(\bold x_5, \bold x_6) = (F_{1,2}^{(1)}(\bold x_5, \bold x_6))^2 F_{22,20}^{(2)}(\bold x_5, \bold x_6)\ , 
\eeq
where the subscript pairs denote the bi-degrees. One may then go through all of the codimension-two points obtained by intersecting $F^{(1)}$ and $F^{(2)}$, and analyze the vanishing orders of $f_J$, $g_J$, and $\Delta_J$ there. It then turns out that at exactly two of those codimension-two points they vanish to order $4$, $6$, and $12$, respectively, and these two points are exactly where the fiber goes non-flat.  We take this as a non-trivial piece of evidence that vanishing of $f$, $g$, and $\Delta$ to those orders is intimately related to the non-flat fibers. 

The above results are somewhat to be expected given results in both the physics \cite{Braun:2013nqa,Morrison:1996na,Seiberg:1996vs,Morrison:1996pp,Candelas:2000nc,Braun:2011ux} and mathematics \cite{RMiranda,MarkG} literatures. It is well known that non-flat fibers are dangerous in that they lead to SCFTs when they are blown down to obtain the associated Weierstrass model. So called ``$(4,6,12)$" points are also well known to be associated with SCFTs, dual to heterotic small instantons, and so it is not surprising that this is what we obtain. Note that, while normally in the physics literature $(4,6,12)$ points are dealt with by blowing up the base, here we see that resolving via a non-flat fiber may also be an option. Indeed  it is known that non-flat fibers can always be birationally related to a flat fibration, by going to the Weierstrass model and then blowing up the base \cite{RMiranda,MarkG}. Thus if one can blow up to a non-flat fiber to resolve some given $(4,6,12)$ points, then removing them by blowing up the base will also be an option.

\section{Explicit Rational Maps}
\subsection{The rational map in Eq.~\eqref{X3-1-sm}}\label{erm-X3-1}
The coefficients for the respective basis monomials in the expression~\eqref{X3-1-sm} are given as
{\small\bea \nn
c_{A_{1,0}}&=& \{374645428, 2459924454, 6041201902, 6757197031, 3506712251, 580810171, 924301148, 8199431465, \\ \nn
&&
25264031186, 36372427880, 22779823774, 4908357591, 1452188771, 13033199001, 40135221388, \\ \nn
&& 66573888839, 54108574906, 15515879412, 1248897453,  12094877912,  34197169604, 54193633181, \\ \nn
&&55266938999, 22696202323, 215837430, 6368649137, 15892630074, 22118006553, 22024966369, \\ \nn
&&15692070369, 237121534, -325372751, 6378759403, 1293660759, 3196094763, 3924307629\}\ , \\ \nn
c_{A_{1,1}}&=& \{-734944740, -4326339166, -11130294298, -13907839771, -9009394083, -2347190168,  \\ \nn
&&-1675319136, -14776875073, -46120017979, -63136826781, -36725168560, -7668395811,\\ \nn &&-2113655527, -18322779267, -72678726227, -110406161553, -65489903673, -10031189305, \\ \nn 
&& -1349647374, -11165685192, -53929588927, -95298609096, -59057362305, -9086250737, \\ \nn
&& -625397263, -1377088015, -21506754436, -41349413418, -24925221695, -5744117643, \\ \nn 
&& -270489017, 1411511282, -5280035471, -6318376072, -4098094623, -1500696270 
\}\ , \\ \nn 
c_{A_{2,0}}&=& \{1084, 723, -1357, 446, 1061, -1858, 2295, -4617, 1014\}\ , \\ \nn
c_{A_{2,1}}&=& \{1636, 7255, 4564, 128, 11845, 12591, -166, 3943, 6549\}\ , \\ \nn 
c_{A_{2,2}}&=& \{-2410, -6655, -6728, -2985, -15122, -8890, -1337, -6671, -3039\}\ , \nn
\eea}
where the coefficients are listed in the lexicographic order of the monomial exponents described in Section \ref{eg2}.

\subsection{The rational map in Eq.~\eqref{X3-2-sm}}\label{erm-X3-2}
The coefficients for the respective basis monomials in the expression~\eqref{X3-2-sm} are given as
{\small\bea \nn
c_{A_{1,0}}&=& \{2221064631, 11917255459, 28570974597, 16495328929, 8210461686, 7704389866, \\ \nn
&&
-49864858414, -52760626002, 8036361006, -15034898100, 6923938379, 37655021444 \} \ , \\ \nn
c_{A_{1,1}}&=& \{-2813621667, -13737694558, -23649388867, -13410639788, -10885604364, -16063185502, \\ \nn
&&
34747606348, 35848540654, -10733842857, 14828715552, 4172982765, -22394056833\} \ , \\ \nn
c_{A_{2,0}}&=& \{32876672958, 151940458372, 197121202830, 109428884752, 127447660413, 193402090199, \\ \nn 
&& 
-438295120157, -348851016375, 125317972440, -68136505562, 330460624192, 602425176324, \\ \nn
&&
-2417059116, 78061046172, -189976587911, -401928611774\}\ , \\ \nn
c_{A_{2,1}}&=& \{27539562618, 155918941832, 429523158738, 249442739852, 64390426983, -104328590424, \\ \nn
&&
-1019425293349, -969130178938, -39468341094, -438573483026, 455036821602, 911597522988, \\ \nn
&&
-134257162533, 161878775784, 112484337279, -154768837497\}\ , \\ \nn 
c_{A_{3,0}}&=& \{7357, 26839, 10114, -49331\}\ , \\ \nn
c_{A_{3,1}}&=& \{537, 15147, -1059, 1817\}\ , \\ \nn
c_{A_{3,2}}&=& \{-15452, -30676, -29780, 39679\}\ , \\ \nn
c_{A_{4,0}}&=& \{26125015333068, 14093705690638, -253407065151880, -1392929418892058, -969784333605016, \\ \nn
&&
44111282567346, -403354228064373, 154876615769747, 5112562529762937, 4813145357835159, \\ \nn
&&
-171100860164100, -616912937629296, 868417042593585, -6973432524385874, -8327868925909157, \\ \nn
&&
-436942930362768, 56095959480934, -1955852891330594, 3018949340241019, 5939379450686520, \\ \nn
&&
-233507718098055, -349570577705856, 1272688939129888, 369255734060038, -1443023739731898\}\ , \\ \nn
c_{A_{4,1}}&=& \{52147316163654, 182788360653752, 31326182511870, -568882677708652, -448279427660016, \\ \nn
&& 
196582084116597, 18967518423364, -1149983213111555, 512164610746422, 1360945691525348, \\ \nn
&&
196592267852520, -250589375084200, 1437953867180930, 1652250053613618, -567580289130554,  \\ \nn
&& 
32425178124081, 442219027542972, -56965838050663, -1358907015270799, -779145891064596, \\ \nn
&&
46248190501128, 118523292993366, -333807482278694, -124559271657028, 390617948985922 \} \ , \\ \nn
c_{A_{4,2}}&=& \{-51721845976638, -211044216465212, -185467081673182, 770647194632056, 613395005398400, \\ \nn
&&
 -85575733018929, 267990300915931, 868293730827257, -1937085042649463, -2902140503357580, \\ \nn 
 && 
 217465884106170, 270811223476316, -3411572449024905, 364374416960336, 3657090237062633, \\ \nn
 &&
 361920509039658, -1037318384139832, 2066163139747743, 1800958612708262, -1942857053473884, \\ \nn
&&
-61786399417851, 407958330780186, -684815876214366, -858431195253898, 730146388419428 \} \ , \\ \nn
c_{A_{4,3}}&=& \{-10832442228972, -816963082358, 131545420796888, 537953669498482, 370184917765880, \\ \nn
&&
-99402081447330, -80480182341872, 20890954975173, -916841939725634, -1003292183336129, \\ \nn
&&
-218570855874462, 341593139545558, 1176891692757061, 661234289887252, 235975445849429, \\ \nn
&&
-46147618216851, 487317156800170, -1349531840686582, -1215954798086102, 875569510180860, \\ \nn
&&
175304685193347, -456024004562430, 331856671930056, 689389343218620, -468384360246228 \} \ , 
\eea}
where the coefficients are listed in the lexicographic order of the monomial exponents described in Section \ref{eg2}. 

\subsection{The rational map in Eq.~\eqref{X3-3-sm}}\label{erm-X3-3}
The coefficients for the respective basis monomials in the expression~\eqref{X3-3-sm} are given as
{\small\bea \nn  
c_{A_{1,0}}&=& \{-651897 \} \ , \\ \nn
c_{A_{1,1}}&=& \{61934\} \ , \\ \nn
c_{A_{1,2}}&=& \{548944\} \ , \\ \nn 
c_{A_{2,0}}&=& \{21105788506648, -11775044280804, 6163099181603, 29703669175731, -14775894485118, \\ \nn
&&
-9206484163372, 9487229871540, -4625233450212, -7147022545227
\}\ , \\ \nn
c_{A_{2,1}}&=& \{
-4435876022168, 1454959565556, -2028437292847, -6978840453732, 1754250504522, 421432378118, \\ \nn
&&
-2379663273177, 466722799641, 457564831947
\}\ , \\ \nn 
c_{A_{2,2}}&=&\{
-14587835867512, 8657245681028, -3886990601851, -20156267237980, 10899890597746, \\ \nn
&&
7132962137614, -6361220860757, 3453847947797, 5471027603303
\}\ , \\ \nn
c_{A_{3,0}}&=& \{1836343820, -1284971884, 348994373, 2279657494, -1578310010, -1230865198, 611200657, \\ \nn
&&
-492599473, -835170489\}\ , \\ \nn
c_{A_{3,1}}&=& \{-347211138, -689099572, -736001231, -1590588657, -718071560, -1330998724, -950616680, \\ \nn
&&
 -206565160, -573918245\}\ , \\ \nn
c_{A_{3,2}}&=&\{798897195, -82629092, 494287724, 1618408765, -182328463, 276360492, 726380168, -72433064, \\ \nn
&&
-4829618\} \ , \\ \nn
c_{A_{3,3}}&=&\{-1500914105, 1290270968, -112711221, -1457099424, 1503524551, 1448568784, -177947468,\\ \nn
&&
 443979836, 821860735\}\ , \\ \nn
c_{A_{4,0}}&=& \{67\}\ , \\ \nn
c_{A_{4,1}}&=& \{4\} \ , \\ \nn
c_{A_{5,0}}&=& \{-92\} \ , \\ \nn
c_{A_{5,1}}&=& \{247 \} \ , 
\eea}
where the coefficients are listed in the lexicographic order of the monomial exponents described in Section \ref{eg2}. 

\end{appendix}


\end{document}